\documentclass[a4paper,11pt]{article}
\pdfoutput=1 

\usepackage{jcappub} 
\usepackage{amsmath}
\usepackage{amssymb} 
\usepackage{graphicx}
\usepackage{enumerate}
\usepackage{color}
\usepackage{tensor}
\usepackage{mathrsfs}
\usepackage{hyperref}
\usepackage[normalem]{ulem}
\usepackage[dvipsnames]{xcolor}
\usepackage{accents}
\usepackage{scalerel}
\usepackage{bm}
\usepackage{comment}
\DeclareMathOperator{\diag}{diag}

\makeatletter
\newcommand*{\leqdef}{\mathrel{\rlap{%
			\raisebox{0.25ex}{$\m@th\cdot$}}%
		\raisebox{-0.25ex}{$\m@th\cdot$}}%
	=}
\newcommand*{\reqdef}{=\mathrel{\rlap{%
			\raisebox{0.25ex}{$\m@th\cdot$}}%
		\raisebox{-0.25ex}{$\m@th\cdot$}}
}
\makeatother

\newcommand{\lambdabar}{{\mkern0.75mu\mathchar '26\mkern -9.75mu\lambda}}

\title{Influence of gravitational waves upon light. Part I. Null geodesics, radar distance and frequency shift}

\author[a]{Jo\~ao C. Lobato,}

\author[a]{Isabela S. Matos,}

\author[a]{Lucas T. Santana,}

\author[a, b]{Ribamar R. R. Reis}

\author[a]{and Maur\'\i cio O. Calv\~ao}

\affiliation[a]{Universidade Federal do Rio de Janeiro,
	Instituto de F\'\i sica, \\
	CEP 21941-972 Rio de Janeiro, RJ, Brazil}

\affiliation[b]{Universidade Federal do Rio de Janeiro, Observat\'orio do Valongo, 
	\\CEP 20080-090 Rio de Janeiro, RJ, Brazil}

\emailAdd{jcavlobato@if.ufrj.br}
\emailAdd{isa@if.ufrj.br}
\emailAdd{lts@if.ufrj.br}
\emailAdd{ribamar@if.ufrj.br}
\emailAdd{orca@if.ufrj.br}

\abstract{We explore different facets of the action of linearized gravitational waves in Minkowski spacetime background upon light, under the electromagnetic geometrical optics limit, covering the main aspects: light trajectory perturbations, radar distance and light frequency shift. For this purpose, we consider observers comoving with the transverse traceless gauge coordinates. We compute the parametrized null geodesics exchanged between two of these observers, presenting explicitly the constants of motion as functions of observables, determining therefrom both the radar distance between the observers and the electromagnetic round-trip frequency shift caused by the gravitational wave. Also, a comparison is made between these results and what one would obtain by using a frequently adopted hybrid model in which the spatial trajectory of light is unchanged. Finally, we revisit and provide an explanation, resorting to the constancy of the phase along a light ray, to the fundamental puzzling question of how one is able to detect gravitational waves by means of interferometry if both light wavelength and detector arms are stretched.}


\begin{document}
	
\maketitle
\flushbottom
\section{Introduction}
\label{sec:intro}

The first direct detection of gravitational waves (GW) by interferometric experiments \cite{Abbott2016b} opened a new realm of investigation for our Universe. Although the basic concepts regarding interferometry on relativistic investigations are well known since the Michelson-Morley experiment \cite{Michelson1881, Michelson1887}, the discussion of that procedure in the context of general relativity or its modifications deserves special attention. When dealing with this problem, a first attempt is to consider only the interaction of gravity with massive particles, particularly those determining the extremities of the interferometer arms. If this picture is valid, the whole situation can be analyzed as in flat spacetime, with the addition of gravity only as the agent causing the anisotropic stretch of the arms, which furthermore changes light optical paths, providing a non-trivial interference pattern at the end. However, under this perspective, another relativistic aspect which could be relevant is neglected, namely, light's interaction with gravity. 

The metric of spacetime selects the possible null 4-dimensional rays along which light particles travel when the geometrical optics limit of Maxwell's equations is assumed. It is natural, then, to wonder if considering such interaction in all its possible facets could bring new elements and corrections to the detection process of GWs. 
On the other hand, since GWs are extremely weak, one could argue that their effect on light cannot be measured in any manner. Nonetheless, we know how interferometry acts as an amplifier of small disturbances and also, as will be discussed later on, that such interaction is already present at linear order in the GW amplitude. 

Several studies have approached this subject in a variety of aspects, such as: (i) deriving the complete null-geodesics in a GW spacetime \cite{Rakhmanov2009, Bini2009, deFelice2010}, (ii) light spatial trajectory perturbations \cite{Finn2009}, (iii) electromagnetic frequency shift \cite{Kaufmann1970, Estabrook1975, Tinto1998, Armstrong2006, Tinto2002} and (iv) electric field evolution along light rays \cite{Santana2020}. In this series of two works, we aim at investigating subjects (i), (ii) and (iii) (theme of this first work) in order to relate them with (iv) when applied to an idealized interferometry experiment (theme of the second work \cite{Lobato2021Part2}, hereafter referred to as L2). The change in luminous spatial path could alter the round-trip travel time of light (or equivalently, the radar length of the arm) and a frequency shift could, in principle, influence the phase and intensity of the electromagnetic field. Even though our ultimate concern is with the interferometric procedure, most of the results obtained in this first paper are valid in a broader picture, where two observers comoving with the transverse traceless (TT) gauge coordinates in a GW spacetime with flat background exchange light rays with each other. We point out that, differently from most of the above mentioned works, here all final quantities are explicitly calculated, not in terms of arbitrary constants of motion, but of known parameters (observables) through the imposition of what we shall call mixed conditions (cf. eq.~(\ref{mixed_conditions})).

In section \ref{sec:models} we carefully describe our family of models, formalizing our discussion, and comment on an alternative (hybrid) one that is frequently and implicitly adopted in many works \cite{Baskaran2004, Rakhmanov2008}, in which perturbations of light spatial trajectories are disregarded. In section \ref{sec:RD_definition}, for a generic spacetime, we define the radar distance and call special attention to the mixed conditions that will be imposed here and through which it is uniquely characterized in terms of known quantities. In section \ref{sec:RD_unperturbed_perturbed}, we compute the radar distance within the two family of models and arrive at the conclusion that they provide the same expression; we spell out the reasons for this coincidence to hold, complementing Finn's illuminating article \cite{Finn2009}. In section \ref{sec:Doppler_effect} we present the parametrized null geodesic arcs, derive the expression for the round-trip frequency shift, interpret it as a Doppler effect and use one of the fundamental laws of electromagnetic geometrical optics to tackle the common conundrum regarding the possibility of detecting GWs when both arm and light's wavelength are stretched \cite{Faraoni2007, Saulson1997}. In section \ref{sec:conclusion} we discuss our main results and point out further directions.

Our signature is $+2$ and we set $c = 1$. We use Latin letters at the middle of the alphabet ($i$, $j$, $k$, $l$, $m$, $n$) to denote spatial indices, and at the end of the alphabet ($t$, $x$, $u$, $v$, $y$, $z$) to denote actual coordinate indices. Greek component indices denote either a spatial or a temporal index. As concerns the concepts of instantaneous observer, observer, reference frame, photon and light signal, we consistently adhere to \cite{Sachs1977}; in particular, a reference frame is conceived as a continuous system and, consequently, its motion can be described just as in the Newtonian kinematics of an ordinary fluid (see, for example, \cite{Ellis1971}).
\section{Models}
\label{sec:models}
The models we will consider throughout this article to describe light rays propagating in vacuum in a gravitational wave (GW) spacetime, unless explicitly stated, constitute a two-parameter family defined by the 6-tuple
\begin{equation}
\mathbb{M}_{(\bm{\epsilon})} \leqdef \left(\,{\mathcal N}, \,\varphi, \,g_{(\bm{\epsilon})}, \,u, \,\xi_{(\bm{\epsilon})}, \,\bm{P}\,\right)\,, \label{consistent_models}
\end{equation}
where
\begin{itemize}
	\item $\bm{\epsilon} \leqdef (\epsilon_+,\epsilon_\times)$ is a pair of independent parameters that will be considered up to linear order in all calculations.
	\item ${\mathcal N}$ is a bare 4-dimensional differentiable manifold;
	\item $(\mathcal{U}, \varphi)$ is a fixed chart on an open set $\mathcal{U} \subset \mathcal{N}$. Its coordinate functions $x^{\mu}$ are denoted by $(t, x, y, z)$ and will be called TT coordinates;
	\item $g_{(\bm{\epsilon})}$ is an $\bm{\epsilon}$-parametrized family of Lorentzian metric fields on ${\mathcal N}$, TT gauge solutions of the linearized vacuum Einstein equations in a Minkowski background. Its components in the $\varphi$ coordinate basis, are:
	\begin{align}
	g_{(\bm{\epsilon})\alpha \beta}(t - x) &\leqdef  \eta_{\alpha \beta} + \epsilon_P h^P_{\alpha \beta}(t - x), \label{metric} \\
	\eta_{\alpha\beta} &\leqdef \diag(-1, 1,1,1)\,, \label{Minkowski_matrix}
	\end{align}
	where the two GW polarizations are indexed by $P= +, \times$, for which the usual Einstein summation convention holds, and
	\begin{align}
	h^+_{\alpha \beta}(t-x) & \leqdef h_{+}(t-x)(\delta_{\alpha 3}\delta_{\beta 3} - \delta_{\alpha 2}\delta_{\beta 2})\,, \label{hplus} \\
	h^{\times}_{\alpha \beta}(t-x) & \leqdef -h_{\times}(t-x) (\delta_{\alpha 2} \delta_{\beta 3}+\delta_{\alpha 3} \delta_{\beta 2})\,, \label{hcross}
	\end{align}
	specify the line element of a GW traveling in the $x$ axis
	\begin{align}
	 ds^2 = & - dt^2 + dx^2 + [1 - \epsilon_+ h_{+}(t - x)]dy^2 + [1 +\epsilon_+ h_{+}(t-x)] dz^2 \nonumber \\  
	&  - 2 \epsilon_{\times} h_{\times}(t-x) dy dz\,. \label{line_elem_TT}
	\end{align}
	
	\item $u$ is the 4-velocity field defining the reference frame (congruence of observers, i.e. of timelike future-directed worldlines \cite{Sachs1977}) comoving (adapted) with the TT coordinate system, namely,
	\begin{equation}
	u  \leqdef \frac{\partial_t}{\sqrt{-g_{(\bm{\epsilon})}(\partial_t, \partial_t)}} = \partial_t\,.
	\label{TTframe}
	\end{equation}
	This will be called the TT reference frame and, of course, $x^i =$ const along each of its observers. We anticipate that, for many of our purposes, only a few of the uncountable observers that constitute such a frame will be necessary and experimentally relevant.
	
	\item $\xi_{(\bm{\epsilon})}$ is an affinely parametrized null geodesic curve, function of a real parameter $\vartheta$, associated to the metric (\ref{metric}), that is,
	\begin{align}
	g_{(\bm{\epsilon})}(k_{(\bm{\epsilon})}, k_{(\bm{\epsilon})}) & = 0\,, \label{pert_nullity}\\ 
	\frac{D_{(\bm{\epsilon})}}{d\vartheta} k_{(\bm{\epsilon})} & = 0\,, \label{pert_geod}
	\end{align}
	where
	\begin{align}
	k_{(\bm{\epsilon})} \leqdef \frac{d \xi_{(\bm{\epsilon})} }{d\vartheta}\,, \label{pert_velocity}
	\end{align}
	 and $D_{(\bm{\epsilon})}/d\vartheta$ is the directional absolute derivative related to the metric $g_{(\bm{\epsilon})}$ along the light ray $\xi_{(\bm{\epsilon})}$.
	
	\item $\bm{P}$ is a specific, physically well motivated, list of mixed (initial and boundary) data (parameters) we shall impose on $\xi_{(\bm{\epsilon})}$. Which exact conditions are associated to them and why they are necessary is discussed in section \ref{sec:RD_definition}. 
\end{itemize}

It is immediate to note from the description above that the unperturbed model $\mathbb{M}_{(\bm{0})}$ consists of the Minkowski spacetime as represented through a pseudo-Cartesian coordinate system and its comoving inertial reference frame, whose observers may send light rays to each other via the curves $\xi_{(\bm{0})}$. Every quantity related to this zero-th order model, i.e. those evaluated at $\bm{\epsilon} = \bm{0}$ everywhere, will be  written with a subscript $(\bm{0})$ juxtaposed to its kernel symbol.  In all forthcoming sections, $\bm{\epsilon}$ dependencies in the arguments of functions will be omitted and whenever $\bm{\epsilon}$ is arbitrary so will the sub-indices $(\bm{\epsilon})$.

Moreover, in subsection \ref{subsec:unperturbed_spatial_trajectories} only, a second family of notorious \emph{hybrid} models will be assumed:
\begin{equation}
\mathbb{M}_{\textrm{hyb}(\bm{\epsilon})}\leqdef \left( \mathcal{N},\,\varphi, \,g_{(\bm{\epsilon})}, \,u, \,\tilde{\xi}_{(\bm{\epsilon)}}, \,\bm{P} \right)\,. \label{inconsis_model}
\end{equation} 
The only difference between this family and that provided by eq.~(\ref{consistent_models}) is simply the curve used to describe light. Here, instead of $\xi_{(\bm{\epsilon})}$, we use the curve $\tilde{\xi}_{(\bm{\epsilon})}$ that satisfies
\begin{equation}
\tilde{\xi}^{i}_{(\bm{\epsilon})} \leqdef \xi_{(\bm{0})}^i \label{xi_tilde}
\end{equation}
and an equation analogous to (\ref{pert_nullity}) for its tangent vectors $\tilde{k}_{(\bm{\epsilon})}$, from which $\tilde{\xi}^t_{(\bm{\epsilon})}$ can be uniquely determined. In other words, $\tilde{\xi}_{(\bm{\epsilon})}$ is a null curve in the current spacetime, with a spatial trajectory coincident with that of the unperturbed curve $\xi_{(\bm{0})}$. As will become explicit later, this second family of models is, apart from very particular circumstances, inconsistent for the description of light propagation since $\tilde{\xi}_{(\bm{\epsilon})}$ is not a geodesic for the metric $g_{(\bm{\epsilon})}$. Nevertheless, eq.~(\ref{inconsis_model}) will be considered to expose the procedure one would make if light's spatial trajectory perturbations due to its interaction with GWs were neglected, an assumption commonly found in the literature \cite{Baskaran2004, Rakhmanov2008}. Table \ref{tab:symbols} is a helpful reference for the notation employed throughout the whole series. 

Last, we emphasize that the metric (\ref{metric}) is a solution of the Einstein's field equations in vacuum, so another possible aspect of the interaction between GWs and electromagnetic waves, namely, the effect of light energy-momentum tensor as a source to the curvature of spacetime will not be considered in this work, that is, light will be held as a test field only (cf. \cite{Schneiter2018}).
\begin{table*}
	\label{tab:symbols}	
	\begin{tabular}{|c|l|c|}	
		\hline
		\textbf{Symbol} & \hspace{3.5cm} \bfseries{Description} & \textbf{Reference(s)} \\
		\hline 
		$\mathbb{M}_{(\bm{\epsilon})}$ & Real perturbed models & \eqref{consistent_models} \\
		$\mathbb{M}_{\textrm{hyb}(\bm{\epsilon})}$ & Hybrid models & \eqref{inconsis_model} \\
		$\bm{\epsilon}, \epsilon_P$ & Perturbation control parameters & - \\
		$\mathcal{N}$ & Base 4-dimensional manifold & - \\
		$(\mathcal{U},\varphi)$ & Transverse traceless (TT) chart whose coordinates are $(t,x,y,z)$ & - \\
		$(u,v,y,z)$ & Coordinate functions of a light-like chart  & \eqref{null_coordinates} \\
		$g_{(\bm{\epsilon})}, g$ & Perturbed family of metrics & \eqref{metric}, \eqref{Minkowski_matrix} \\
		$h^P_{\alpha \beta}$ & Components of the first-order perturbation of the metric & \eqref{metric}--\eqref{hcross} \\
		$h_P$ & Non-vanishing components of $h^P_{\alpha \beta}$ & \eqref{hplus}--\eqref{line_elem_TT} \\
		$u$ & TT adapted frame & \eqref{TTframe} \\
		$\tau$ & Proper time of TT observers & - \\
		$\mathcal{S}$ & Source observer & Figures~\ref{fig:radar_distance}, \ref{fig:radar_distance_3_types_of_curve} \\
		$\mathcal{M}$ & Mirror observer & Figures~\ref{fig:radar_distance}, \ref{fig:radar_distance_3_types_of_curve} \\
		$\mathcal{E}$ & Emission event & Figures~\ref{fig:radar_distance}, \ref{fig:radar_distance_3_types_of_curve} \\
		$\mathcal{R}$ & Reflection event & Figures~\ref{fig:radar_distance}, \ref{fig:radar_distance_3_types_of_curve} \\
		$\mathcal{D}$ & Detection event & Figures~\ref{fig:radar_distance}, \ref{fig:radar_distance_3_types_of_curve} \\
		$\xi_{(\bm{\epsilon})}, \xi$ & Null affinely parametrized geodesics (arcs of) & \eqref{pert_velocity}--\eqref{pert_geod} \\
		$\vartheta$ & Affine parameter of null geodesics $\xi_{(\bm{\epsilon})}$ & \eqref{pert_velocity}--\eqref{pert_geod} \\
		$k_{(\bm{\epsilon})}, k$ & Wave number 4-vector of $\xi$ & \eqref{pert_velocity} \\
		$\omega_{\textrm{e}}$ & Electromagnetic wave angular frequency & - \\
		$n$ & Unit vector along the spatial projection (onto the rest space) of $k_{(\bm{\epsilon})}$ & Figure \ref{fig:arm_direction} \\
		$\bm{P}$ & Mixed data (parameters) for curves $\xi_{(\bm{\epsilon})}$ and $\tilde{\xi}$ & \eqref{mixed_conditions} \\
		$\tilde{\xi}_{(\bm{\epsilon})}, \tilde\xi$ &  Null, non-geodesic rays with unperturbed spatial trajectories & \eqref{xi_tilde} \\
		$\tilde{k}_{(\bm{\epsilon})}$, $\tilde{k}$ & Wave number 4-vector of $\tilde{\xi}$ & -\\
		$D_{(\bm{\epsilon})}/d\vartheta$ & Absolute directional derivative along $k_{(\bm{\epsilon})}$ & - \\
		$D_{R,\xi}$, $D_{R,\tilde{\xi}}$ & Radar distance between $\mathcal{S}$ and $\mathcal{M}$ using, respectively, $\xi$ and $\tilde{\xi}$& \eqref{D_R_general},\eqref{D_R_int} \\
		$\Delta\ell$ & Unperturbed radar distance & \eqref{delta_l}\\
		$\delta$ & Constant of geodesic motion associated to $\partial/\partial v$ Killing vector & \eqref{kv} \\ 
		$\alpha$ & Constant of geodesic motion associated to $\partial/\partial y$ Killing vector & \eqref{ky} \\
		$\beta $ & Constant of geodesic motion associated to $\partial/\partial z$ Killing vector & \eqref{kz} \\
		$A$ & Ratio $\alpha/\delta$& \eqref{eq:constant_of_motion_A} \\
		$B$ & Ratio $\beta/\delta$ &\eqref{eq:constant_of_motion_B} \\
		$m_P$ & Integrated GW amplitude along a light ray as a function of $\vartheta$ & \eqref{n1}, \eqref{w} \\
		$M_P$ & Integrated GW amplitude as a function of time& \eqref{M_P_1}, \eqref{M_P_2} \\
		$\theta$ & Zenithal angle with respect to the GW propagation direction $\partial/\partial x$ & Figure \ref{fig:arm_direction} \\
		$\phi$ & Azimuthal angle & Figure \ref{fig:arm_direction} \\
		\hline	
	\end{tabular}
	\caption{\sc{Symbols used in this series of articles.}}
\end{table*}
\section{Radar distance definition and mixed conditions} 
\label{sec:RD_definition}

\begin{figure}
	\centering
	\includegraphics[scale=0.2]{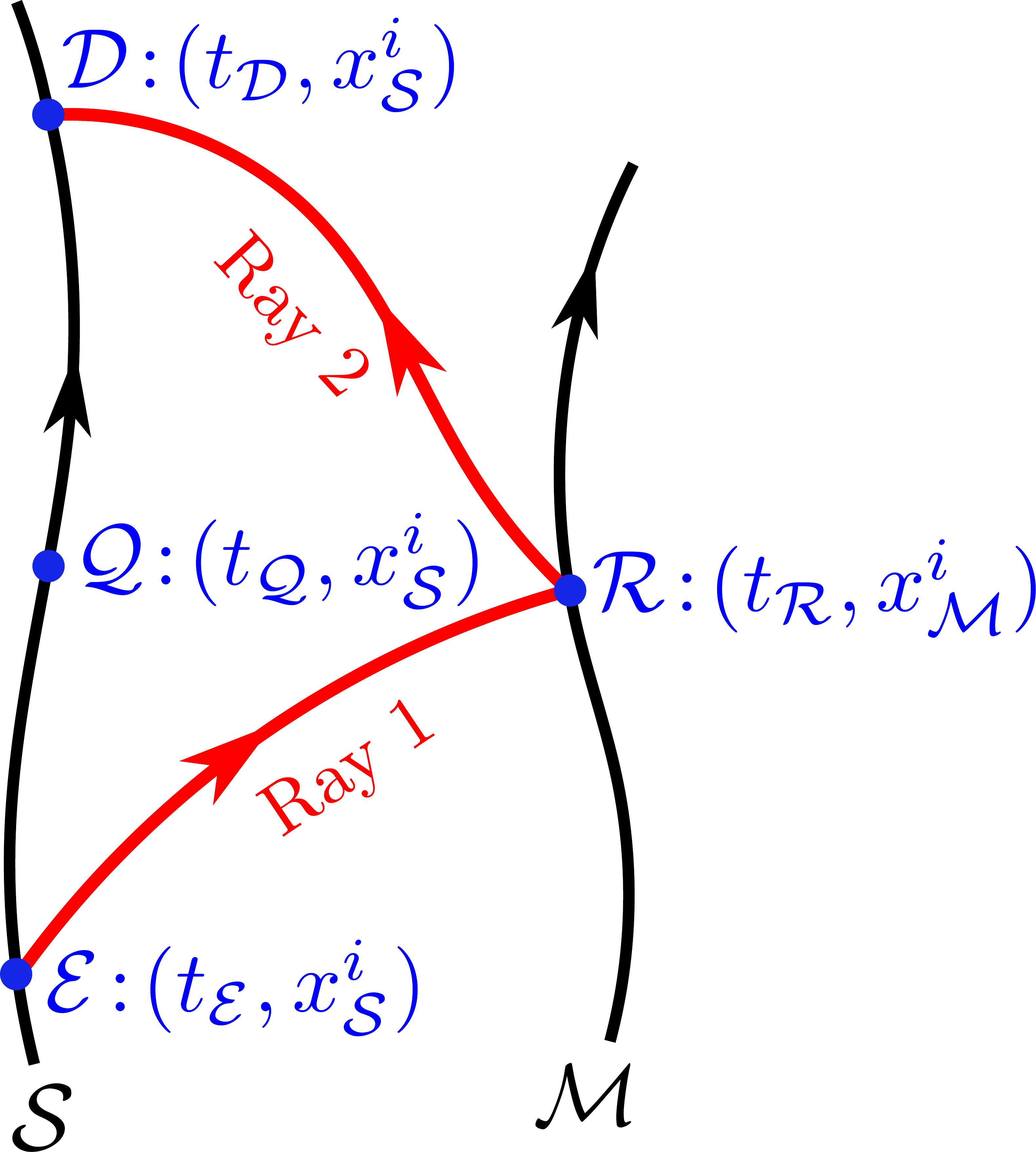} 
	\caption{Observer $\mathcal{S}$ ascribes a radar distance, at its event $\mathcal{Q}$, corresponding to its proper time $\tau_{\mathcal{Q}}$, to observer $\mathcal{M}$, via null geodesic rays 1 and 2.}
	\label{fig:radar_distance}
\end{figure}

We start by defining the radar distance between two observers, $\mathcal{S}$ (source) and $\mathcal{M}$ (mirror), with the aid of figure~\ref{fig:radar_distance}. Observer $\mathcal{S}$ emits a photon from event $\mathcal{E}$ with proper-time $\tau_{\mathcal{E}}$, which travels along ray 1 towards observer $\mathcal{M}$. There, at event $\mathcal{R}$, it is reflected and travels back along ray 2, whereupon it is detected at the event $\mathcal{D}$, with proper time $\tau_{\mathcal{D}}$.

The radar distance $D_R$ that $\mathcal{S}$ assigns to $\mathcal{M}$ at the ``mid-point'' event $\mathcal{Q}\in\mathcal{S}$, defined to have proper time
\begin{equation}
\tau_{\mathcal{Q}} \leqdef \frac{\tau_{\mathcal{E}} + \tau_{\mathcal{D}}}{2}\,, \label{tQ}
\end{equation}
is
\begin{equation}
D_R(\mathcal{S}, \mathcal{M}, \tau_{\mathcal{Q}}) \leqdef \frac{\tau_{\mathcal{D}} - \tau_{\mathcal{E}}}{2}\,.  \label{D_R_general}
\end{equation}
As defined, such a distance is a purely geometric scalar, with all the needed information being measured by the single observer $\mathcal{S}$. Furthermore it does not rely on a given extended reference frame, as its construction assumes only two observers.

Restricting ourselves now to the models discussed in section \ref{sec:models}, observers $\mathcal{S}$ and $\mathcal{M}$ will belong to the TT frame given by eq.~(\ref{TTframe}). Since from eq.~(\ref{line_elem_TT}) $g_{tt} = -1$, the time-like coordinate $t$ along an adapted observer is equal to its proper time $\tau$, and we may also write:
\begin{equation}
D_R(\mathcal{S}, \mathcal{M}, t_{\mathcal{Q}}) = \frac{t_{\mathcal{D}} - t_{\mathcal{E}}}{2}\,.
\end{equation}   

The procedure described above to determine $D_R$ will only be successful once certain boundary conditions are imposed for rays 1 and 2, so that one assures light reaches observer $\mathcal{M}$ and gets back to $\mathcal{S}$. The coordinate representations of rays 1 and 2 will be denoted by $\xi^{\alpha}_{|j}$, where $j = 1,2$ indexes each ray. Then, for each instant $\xi^t_{|1}(0) \reqdef t_{\mathcal{E}}$ of emission, the partial boundary conditions for ray 1 are:
\begin{align}
&\xi^i_{|1}(0) = x^i_{\mathcal{S}}\,, \label{boundE}\\
&\exists \; \vartheta_{\mathcal{R}} >0 \; | \; \xi^i_{|1}(\vartheta_{\mathcal{R}}) = x^i_{\mathcal{M}}\,, \label{boundR1}
\end{align}
where the fixed spatial coordinates of $\mathcal{S}$ and $\mathcal{M}$ are denoted by $x^i_{\mathcal{S}}$ and $x^i_{\mathcal{M}}$, respectively. These conditions ensure that ray 1 connects events $\mathcal{E}$ and $\mathcal{R}$. For ray 2:
\begin{align}
&\xi^i_{|2}(0) = x^i_{\mathcal{M}}\,, \label{boundR2} \\ 
&\exists \; \vartheta_{\mathcal{D}} >0 \; | \; \xi^i_{|2}(\vartheta_{\mathcal{D}}) = x^i_{\mathcal{S}}\,, \label{boundD}
\end{align}
which guarantees that ray 2 connects events $\mathcal{R}$ and $\mathcal{D}$.

For each emission time $t_{\mathcal{E}}$, these conditions allow one to obtain the radar distance in terms of known quantities. Moreover, they will select, from the family of all possible null geodesics, unique \emph{parametrized} arcs for rays 1 and 2, provided that an initial value ${\omega_{\textrm{e}}}_{\mathcal{E}}$ for the frequency of light $\omega_{\textrm{e}|1}(\vartheta)$ is given additionally:
\begin{equation}
	{\omega_{\textrm{e}}}_{\mathcal{E}} = \omega_{\textrm{e}|1}(0) = - k_{|1}^{\mu}(0)u_{\mu}(\xi_{|1}(0)) = k_{|1}^t(0)\,. \label{initfreq}
\end{equation}
Of course, the radar distance must be independent of this choice of frequency (achromaticity); in other words, it only depends on the light signal equivalence class, not its specific photon representative \cite{Sachs1977}. Note that, assuming light is emitted from a laser attached to $\mathcal{S}$, since the inner workings of such a device are solely determined by its atomic structure, the initial value of its frequency is not disturbed by the feeble GW, being then independent of $\bm{\epsilon}$. Such a frequency will however evolve throughout the light ray 1 in a non-trivial fashion due to the GW. This will be explored in section \ref{sec:Doppler_effect}. Finally, to connect the frequency of ray 1 at event $\mathcal{R}$ with the initial frequency of ray 2, we assume there occurs a reflection by a mirror at rest on the TT frame, and thus,
\begin{equation}
\omega_{\textrm{e}|1}(\vartheta_{\mathcal{R}}) = \omega_{\textrm{e}|2}(0).\label{continuity}
\end{equation} 

The collection of conditions given by eqs.~(\ref{boundE} -- \ref{continuity}) will be called \emph{mixed conditions}, and the parameters appearing therein constitute the ingredient
\begin{equation}
	\bm{P} \leqdef \left(\,x^0_{\mathcal E}, x^i_{\mathcal S}, x^i_{\mathcal M}, {\omega_{\textrm{e}}}_{\mathcal{E}}\,\right) \label{mixed_conditions}
\end{equation}
of the models (\ref{consistent_models}). Analogous mixed conditions must be imposed on $\tilde{\xi}$. 

\section{Light spatial trajectories and their radar distance coincidence} 
\label{sec:RD_unperturbed_perturbed}

In this section, inspired by the discussions appearing in \cite{Rakhmanov2008, Rakhmanov2009, Finn2009}, we investigate if the spatial trajectory perturbations on the null geodesics due to the GW give rise to linear order corrections on the radar distance between two TT observers.

\begin{figure}
	\centering
	\includegraphics[scale=0.2]{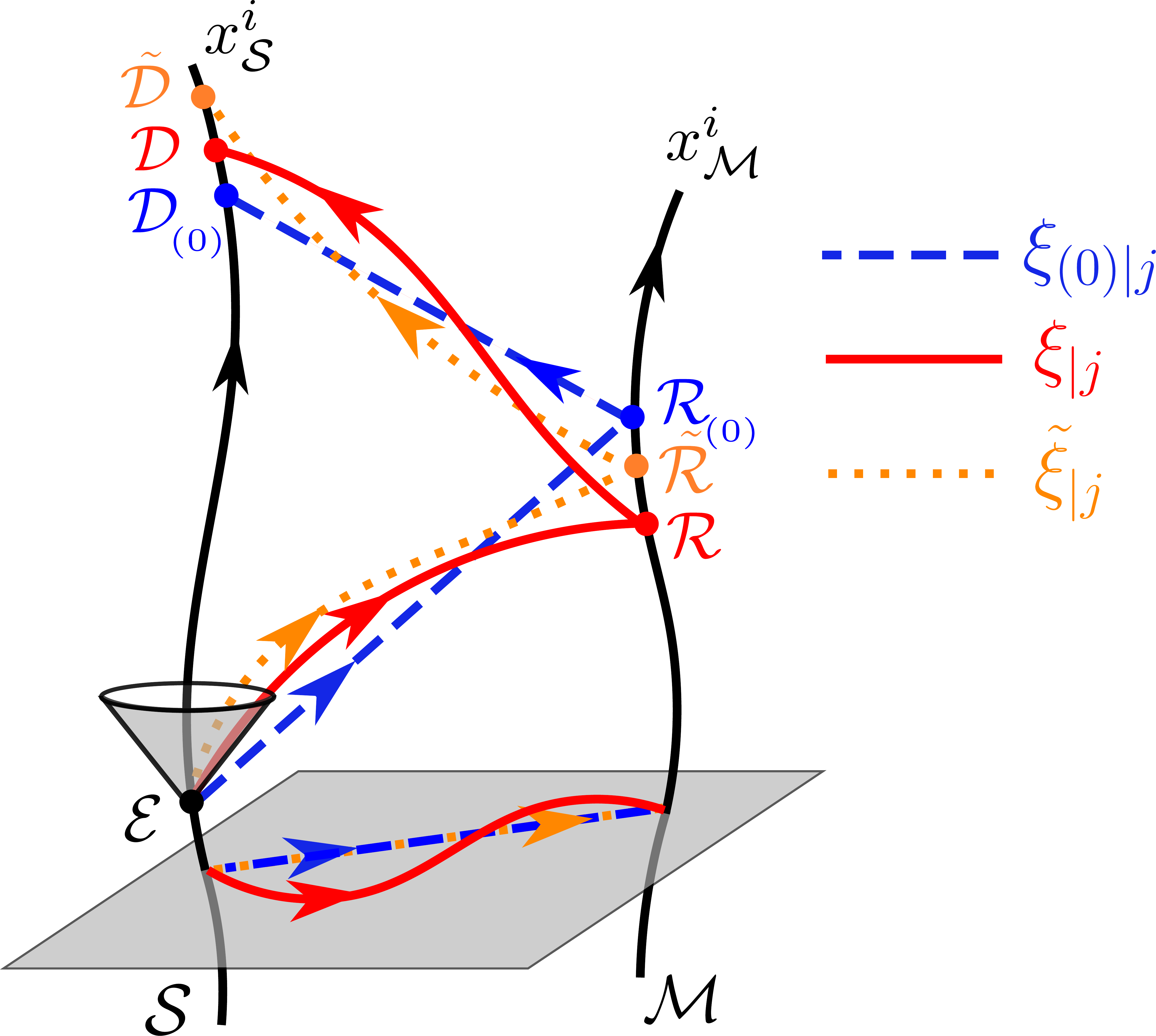}  
	\caption{Minkowski light rays (blue dashed), $\xi_{(\bm{0})|j}$, and perturbed null geodesic rays (red solid), $\xi_{|j}$, connecting two observers. The outgoing arcs ($\xi_{(\bm{0})|1}$ and $\xi_{|1}$, respectively) are constrained to leave observer $\mathcal{S}$ at the same event $\mathcal{E}$ and with the same angular frequency $\omega_{\textrm{e}\mathcal{E}}$; therefore, in general, they reach observer $\mathcal{M}$ at distinct events ($\mathcal{R}_{(\bm{0})}$ and $\mathcal{R}$, respectively) and come back to $\mathcal{S}$, along the incoming arcs ($\xi_{(\bm{0})|2}$ and $\xi_{|2}$, respectively) at distinct events as well ($\mathcal{D}_{(\bm{0})}$ and $\mathcal{D}$, respectively). Besides, we also depict the hybrid null, but not geodesic, curve (orange dotted) $\tilde{\xi}$. Blue and orange curves coincide on the common rest space of $\mathcal{S}$ and $\mathcal{M}$. Whether $\mathcal{R}$ and $\mathcal{\tilde{R}}$ ($\mathcal{D}$ and $\mathcal{\tilde{D}}$) coincide is the subject of section \ref{sec:RD_unperturbed_perturbed}.}
	\label{fig:radar_distance_3_types_of_curve}
\end{figure}

For the discussion in this section, we refer to Figure \ref{fig:radar_distance_3_types_of_curve}\,. It shows the configurations of relevant light rays connecting two observers: Minkowski light rays (blue dashed), $\xi_{(\bm{0})|j}$, perturbed null geodesic rays (red solid), $\xi_{|j}$, and the hybrid null, not geodesic, curve (orange dotted) $\tilde{\xi}$. When a GW reaches an interferometer, it changes the radar length of each arm in an anisotropic way, resulting in a non-trivial intensity pattern due to the difference in light travel times in both arms. Should the studied observers $\mathcal{S}$ and $\mathcal{M}$ represent the extremities of the arm of a GW detector, if there is a correction to the radar length due to the perturbation in light's spatial trajectory, one would expect additional corrections to the intensity pattern as well.
In order to verify if spatial perturbations actually induce such corrections, we will calculate the radar distance candidate $D_{R, \tilde{\xi}}$ in subsection \ref{subsec:unperturbed_spatial_trajectories}, using the models (\ref{inconsis_model}) and thus the orange dotted rays $\tilde{\xi}_{|j}$. Then, in subsection \ref{subsec:perturbed_trajectory}\,, the actual radar distance $D_{R, \xi}$ will be calculated within the models (\ref{consistent_models}), and thus using the red solid curves $\xi_{|j}$.  Note that, for both radar distances, we assume the photon to be emitted at the same event $\mathcal{E}$, while events of reflection ($\mathcal{R}$ and $\tilde{\mathcal{R}}$), and reception ($\mathcal{D}$ and $\tilde{\mathcal{D}}$) do not coincide \emph{a priori}.

\subsection{Using unperturbed light spatial trajectories} 
\label{subsec:unperturbed_spatial_trajectories}

Here we systematically employ the hybrid models (\ref{inconsis_model}). In this approach, the computation of the radar distance candidate is done without considering light spatial trajectory perturbations \cite{Rakhmanov2008, Finn2009}, therefore relying on using the curves $\tilde{\xi}$, where
\begin{equation}
\tilde{\xi}^{i}_{|j} (\vartheta) = \xi^{i}_{(\bm{0})|j}(\vartheta) = k^{i}_{(\bm{0})|j} \vartheta + \xi^{i}_{|j}(0)\,. \label{unptraj}
\end{equation}

Since $\tilde{\xi}$ and $\xi_{(\bm{0})}$ obey the partial boundary conditions (\ref{boundE} -- \ref{boundD}), the above equation shows that $\vartheta_{\mathcal{R}_{(\bm{0})}} = \vartheta_{\tilde{\mathcal{R}}}$ and $\vartheta_{\mathcal{D}_{(\bm{0})}} = \vartheta_{\tilde{\mathcal{D}}}$. 

Imposing $ds^2 = 0$ along $\tilde{\xi}_{|1}$ in eq.~($\ref{line_elem_TT}$), solving for $d\tilde{\xi}^t_{|1}/d\vartheta$ and integrating along ray 1:
\begin{equation}
\Delta t_{\mathcal{E}, \tilde{\mathcal{R}}} = \int^{\vartheta_{\mathcal{R}_{(\bm{0})}}}_{0} \sqrt{[\delta_{ij} + \epsilon_P h^P_{ij}(w_{|1}(\vartheta))] k^i_{(\bm{0})|1} k^j_{(\bm{0})|1}} d\vartheta, \label{timeERunp}
\end{equation}
where, for any pair of events $\mathcal{A}$ and $\mathcal{B}$,
\begin{equation}
\Delta t_{\mathcal{A}, \mathcal{B}} \leqdef t_{\mathcal{B}} - t_{\mathcal{A}}\,,
\end{equation}
and
\begin{align}
w_{|j}(\vartheta) &\leqdef \xi^t_{|j}(\vartheta) - \xi^x_{|j}(\vartheta) \nonumber \\ & = \tilde{\xi}^t_{|j}(\vartheta) - \tilde{\xi}^x_{|j}(\vartheta) + \mathcal{O}(\epsilon_P). \label{w}
\end{align}

Note that $\Delta t_{\mathcal{E}, \tilde{\mathcal{R}}}$ takes into account the perturbations on the metric, though not on the light spatial path, so that it is not the same as $\Delta t_{\mathcal{E}, \mathcal{R}_{(\bm{0})}}$. Moreover, the error in evaluating $\epsilon_P h^P_{ij}$ on $\xi(\vartheta)$ or $\tilde{\xi}(\vartheta)$ is of $\mathcal{O}(\epsilon^2_P)$, since $\tilde{\xi}^t_{(\bm{0})} = \xi^t_{(\bm{0})}$, which justifies the second equality of eq.~(\ref{w}) as well.  More broadly, we notice that all quantities directly obtained by all the three curves $\xi, \tilde{\xi}$ and $\xi_{(\bm{0})}$ coincide at zero-th order in $\epsilon_P$, a result that will often come in handy in our discussion.

Imposing the previously mentioned partial boundary conditions (\ref{boundE}) and (\ref{boundR1}) to $\xi_{(\bm{0})|1}$ on eq.~(\ref{unptraj}), choosing $\vartheta_{\mathcal{E}} = 0$ so that $\xi_{(\bm{0})|1}^i(0) = x^i_{\mathcal{S}}$ and remembering that $\eta_{\mu \nu} k^{\mu}_{(\bm{0})}k^{\nu}_{(\bm{0})} = 0$: 
\begin{equation}
k^{i}_{(\bm{0})|1} = \frac{\Delta x^i}{\vartheta_{\mathcal{R}_{(\bm{0})}}}\,, \quad k^t_{(\bm{0})|1} = \frac{\Delta \ell}{\vartheta_{\mathcal{R}_{(\bm{0})}}}\,, \label{k_{(0)}}
\end{equation}
where
\begin{align}
\Delta x^i \leqdef x^i_{\mathcal{M}} - x^i_{\mathcal{S}}\,, \quad \Delta \ell \leqdef \sqrt{\delta_{ij}\Delta x^i \Delta x^j}\,. \label{delta_l}
\end{align}

Expanding (\ref{timeERunp}) and using (\ref{k_{(0)}}):
\begin{equation}
\Delta t_{\mathcal{E}, \tilde{\mathcal{R}}} = \Delta \ell + \frac{\Delta x^i \Delta x^j}{2 \vartheta_{\mathcal{R}} \Delta \ell} \int_{0}^{\vartheta_{\mathcal{R}}} \epsilon_P h^P_{ij} (w_{|1}(\vartheta))  d\vartheta,
\end{equation}
where $\vartheta_{\mathcal{R}_{(\bm{0})}}$ was replaced by $\vartheta_{\mathcal{R}}$ in the second term with an error of $\mathcal{O}(\epsilon^2_P)$. Now, we change the integration variable to 
\begin{align}
w_{(\bm{0})|1}(\vartheta) &= \xi^t_{(\bm{0})|1}(\vartheta) - \xi^x_{(\bm{0})|1}(\vartheta) \nonumber \\ &= (k^t_{(\bm{0})|1}-k^x_{(\bm{0})|1}) \vartheta + t_{\mathcal{E}} - x_{\mathcal{S}}, \label{u_(0)}
\end{align}
such that
\begin{align}
\epsilon_P\, d\vartheta &=  \frac{\epsilon_P}{(k^t_{(\bm{0})|1}-k^x_{(\bm{0})|1})}\, dw_{(\bm{0})|1} = \frac{\epsilon_P\, \vartheta_{\mathcal{R}}}{\Delta \ell -\Delta x}\,dw_{|1}, \label{int_var_change}
\end{align}
Then, defining
\begin{equation}
m_{P|j}(\vartheta) \leqdef \int_{w_{|j}(0)}^{w_{|j}(\vartheta)} h_{P}(w)\, dw , \label{n1}
\end{equation}
and remembering eqs.~(\ref{hplus}) and (\ref{hcross}):
\begin{align}
\Delta t_{\mathcal{E}, \tilde{\mathcal{R}}} = \Delta \ell - \frac{1}{2(\Delta \ell - \Delta x)\Delta \ell}\big[ \epsilon_+(\Delta y^2 - \Delta z^2)\,  m_{+|1}(\vartheta_{\mathcal{R}}) + 2\epsilon_{\times} \Delta y \Delta z\,   m_{\times|1}(\vartheta_{\mathcal{R}}) \big]. \label{tR-tE}
\end{align}

A similar computation for the back-trip gives:
\begin{align}
\Delta t_{\tilde{\mathcal{R}}, \tilde{\mathcal{D}}} =  \Delta \ell - \frac{1}{{2(\Delta \ell + \Delta x)\Delta \ell}} \big[ \epsilon_+ (\Delta y^2 - \Delta z^2)\,  m_{+|2}(\vartheta_{\mathcal{D}}) + 2\epsilon_{\times} \Delta y \Delta z \,  m_{\times|2}(\vartheta_{\mathcal{D}}) \big]\,. \label{tD-tR}
\end{align}

Adding eqs.~(\ref{tR-tE}) and (\ref{tD-tR}), the radar distance candidate between $\mathcal{S}$ and $\mathcal{M}$ at the mid-point event $\tilde{Q}$ is found:
\begin{align}
D_{R,\tilde{\xi}} \leqdef \frac{\Delta t_{\mathcal{E}, \mathcal{\tilde{D}}}}{2} = \Delta \ell \,-\frac{1}{2\Delta \ell} & \left\{ \epsilon_+ \frac{\Delta y^2 - \Delta z^2}{2} \left[\frac{ m_{+|1}(\vartheta_{\mathcal{R}_{(\bm{0})}})}{\Delta \ell - \Delta x} + \frac{ m_{+|2}(\vartheta_{\mathcal{D}_{(\bm{0})}})}{\Delta \ell + \Delta x} \right] \right. \nonumber \\
& \left. \quad +  \epsilon_{\times} \Delta y \Delta z \left[\frac{ m_{\times|1}(\vartheta_{\mathcal{R}_{(\bm{0})}})}{\Delta \ell - \Delta x} +\frac{ m_{\times|2}(\vartheta_{\mathcal{D}_{(\bm{0})}})}{\Delta \ell + \Delta x} \right] \right\} . \label{D_R_int}
\end{align}

Up to this point, we have been following the previously described procedure in which a TT observer sends a single light ray to another at a given known event and receives it back later. This situation, however, can be thought of as an auxiliary scheme to calculate the radar distance $\mathcal{S}$ ascribes to $\mathcal{M}$ as a function of (its proper) time. In other words, as an alternative and useful perspective, for example, in identifying the radar Doppler effect (cf. section \ref{sec:Doppler_effect}), $\mathcal{S}$ can emit (and also detect) photons at all times, probing the distance $D_{R,\tilde{\xi}} (t_{\mathcal{Q}})$ for all instants $t_{\mathcal{Q}}$ in its worldline.

We may rewrite expression (\ref{D_R_int}) to emphasize the dependence of $D_{R,\tilde{\xi}}$ with the time coordinate $t_{\mathcal{Q}}$. It is only necessary to note that,
\begin{align}
\epsilon_P m_{P|1}(\vartheta_{\mathcal{R}}) &= \epsilon_P \int^{t_{\mathcal{R}}- x_{\mathcal{M}}}_{t_{\mathcal{E}} - x_{\mathcal{S}}} h_{P}(w) dw \nonumber \\ &= \epsilon_P \int^{t_{\mathcal{Q}}- x_{\mathcal{M}}}_{t_{\mathcal{Q}} - \Delta\ell - x_{\mathcal{S}}} h_{P}(w) dw \reqdef \epsilon_P M_{P|1} (t_{\mathcal{Q}})\,, \label{M_P_1}
\end{align}
and similarly:
\begin{align}
\epsilon_P m_{P|2}(\vartheta_{\mathcal{D}}) &=  \epsilon_P \int^{t_{\mathcal{D}}- x_{\mathcal{S}}}_{t_{\mathcal{Q}} - x_{\mathcal{M}}} h_{P}(w) dw \nonumber \\ &= \epsilon_P \int^{t_{\mathcal{Q}}+\Delta \ell -x_{\mathcal{S}}}_{t_{\mathcal{Q}}  - x_{\mathcal{M}}} h_{P}(w) dw \reqdef \epsilon_P M_{P|2} (t_{\mathcal{Q}})\,. \label{M_P_2}
\end{align}
Then 
\begin{align}
D_{R,\tilde{\xi}} (t_{\mathcal{Q}}) = \Delta \ell \,- \frac{1}{2\Delta \ell} & \left\{\epsilon_+ \frac{\Delta y^2 - \Delta z^2}{2} \left[\frac{ M_{+|1} (t_{\mathcal{Q}})}{\Delta \ell - \Delta x} +\frac{ M_{+|2}(t_{\mathcal{Q}})}{\Delta \ell + \Delta x} \right]  \right. \nonumber \\
& \hspace{24pt} \left. + \epsilon_{\times} \Delta y \Delta z \left[\frac{ M_{\times|1}(t_{\mathcal{Q}})}{\Delta \ell - \Delta x} + \frac{ M_{\times|2}(t_{\mathcal{Q}})}{\Delta \ell + \Delta x} \right] \right\}. \label{D_R}
\end{align}

As pointed out by \cite{Finn2009}, the procedure adopted in this subsection, although commonly used when discussing an interferometer response to GWs \cite{Baskaran2004, Rakhmanov2008}, seems inconsistent. What, then, is the interpretation of (\ref{D_R})? It gives the radar distance if a photon could still travel along its unperturbed spatial trajectory, obeying eq.~(\ref{unptraj}), even with the presence of GWs, but with the zero-th component of its parametric curve perturbed by them. However, that proves not to be possible, since $\tilde{\xi}_{|j}$, even though satisfying $ds^2=0$, cannot be additionally a geodesic when (\ref{line_elem_TT}) is assumed. In other words, photons cannot travel along their Minkowski spatial trajectories when a GW is present. The only exceptions to this assertion occur for certain very particular values of $\Delta y$  and  $\Delta z$, as discussed in the next section. In the general case, on the other hand, from the solutions for $\xi_{|1}$ obtained in subsection \ref{subsec:perturbed_trajectory}\,,  $\xi^{i}_{|j} \neq \tilde{\xi}^i_{|j}$. Therefore, it is imperative to compute the fully linearly perturbed null geodesics from observer $\mathcal{S}$ to observer $\mathcal{M}$ (i.e. obeying the necessary partial boundary conditions), to rigorously obtain a physically meaningful radar distance. 

The remaining question to be answered is \cite{Rakhmanov2009, Finn2009}: can those trajectory perturbations significantly alter the distance traveled by light? Certainly, the simplifying assumption made in this section forbids a proper analysis of the electromagnetic frequency evolution along the light ray taking the GW into account as we will show in section \ref{sec:Doppler_effect} (cf. also \cite{Kaufmann1970}). Using eqs.~(\ref{pert_nullity}) and (\ref{k_{(0)}}) together with the expansion
\begin{equation}
	k^i(\vartheta) \reqdef k^i_{(\bm{0})}(\vartheta) + \epsilon_P k^{iP}(\vartheta), \label{k_expansion}
\end{equation}
it is easily seen that, for the TT frame,
\begin{equation}
	\omega_{\textrm{e}}(\vartheta) = k^t(\vartheta) = \tilde{\omega}_{\textrm{e}}(\vartheta) +  \frac{\epsilon_P}{2 \Delta \ell}\delta_{ij} \Delta x^j k^{iP}(\vartheta)\,,
\end{equation}
where $\tilde{\omega}_\textrm{e} \leqdef -u_\mu \tilde{k}^\mu$. We can then expect conceptual and practical aspects regarding this interaction that cannot be attainable through the hybrid model from eq.~(\ref{inconsis_model}).
\subsection{Using perturbed light spatial trajectory}
\label{subsec:perturbed_trajectory}
To obtain the actual null geodesics of this GW spacetime, it is useful to change coordinates to
\begin{align}
u \leqdef \frac{t - x}{2}, \quad
v \leqdef \frac{t + x}{2}, \quad y \leqdef y, \quad z \leqdef z. \label{null_coordinates}
\end{align}
In this new coordinate system, the infinitesimal line element becomes
\begin{align}
ds^2 =& \, [1 - \epsilon_+ h_+(2u)] dy^2 + [1 + \epsilon_+ h_+(2u)] dz^2 -\, 2 \epsilon_{\times} h_{\times}(2u) dy dz -\, 4 du dv \,.
\end{align}

Since $v$, $y$ and $z$ are cyclic variables in the line element, there exist three constants of motion along the light rays:
\begin{align}
&\delta \leqdef k_v, \label{kv} \\
&\alpha \leqdef k_y, \label{ky}\\
&\beta\leqdef k_z. \label{kz}
\end{align} 
From these three constants, one may use the inverse metric $g^{\alpha\beta} = \eta^{\alpha\beta} - \epsilon_P h^{\alpha\beta P}$ (where $h^{\alpha\beta P} \leqdef \eta^{\alpha\mu}\eta^{\beta\nu}h^P_{\mu\nu}$) to obtain
\begin{align}
	&k^u = - \frac{\delta}{2} \label{k^u}\,, \\  
	&k^y = (1+h_+) \alpha + h_{\times} \beta \label{k^y}\,, \\ 
	&k^z = (1-h_+) \beta + h_{\times} \alpha\,. \label{k^z}
\end{align}
Integrating $k^{u}$:
\begin{equation}
w(\vartheta) \leqdef \xi^t(\vartheta) - \xi^x(\vartheta) \reqdef 2\xi^u(\vartheta) =  - \vartheta \delta + t_{\mathcal{E}} - x_{\mathcal{S}}. \label{u}
\end{equation}
Using $k^{\mu}k_{\mu} = 0$: 
\begin{equation}
	2 k^v \delta = - [(1+h_+)\alpha^2 + (1-h_+)\beta^2 + 2 \alpha \beta h_{\times}]. \label{k^v}
\end{equation}
Then, for $\delta \neq 0$ \cite{Rakhmanov2009, deFelice2010, Bini2009}, we find the remaining component $k^v$, and the parametric equations associated to the light rays in terms of the constants of motion are determined by integrating (\ref{k^y}), (\ref{k^z}), (\ref{k^v}), together with (\ref{u}). They can be expressed in terms of $m_P$ as:
\begin{align}
\xi^t(\vartheta) = \,& \xi^t(0) - \frac{1}{2} \vartheta\,  (1 + A^2 + B^2)\,\delta \,+ \frac{1}{2} \,\epsilon_+\, m_+ (\vartheta) (A^2 - B^2) + \epsilon_{\times}\, m_{\times}(\vartheta) AB\,, \label{t} \\
\xi^x(\vartheta) = \,& \xi^x(0) + \frac{1}{2} \vartheta\,(1 - A^2 - B^2)\,\delta \,+ \frac{1}{2} \, \epsilon_+\, m_+ (\vartheta)  ( A^2 - B^2) + \, \epsilon_{\times}\, m_{\times}(\vartheta) AB\,, \label{x} \\
\xi^y(\vartheta) = \,& \xi^y(0) + \vartheta A\,\delta  - A\, \epsilon_+\, m_+(\vartheta) - B\, \epsilon_{\times}\, m_{\times}(\vartheta) \,, \label{y} \\
\xi^z(\vartheta) = \,& \xi^z(0) + \vartheta B\,\delta  + B\, \epsilon_+\, m_+(\vartheta) - A\, \epsilon_{\times}\, m_{\times}(\vartheta)\,, \label{z}
\end{align}
where 
\begin{align}
	A\leqdef \alpha/\delta\,,
	\label{eq:constant_of_motion_A} \\
	B\leqdef \beta/\delta\,.
	\label{eq:constant_of_motion_B}
\end{align}

The $m_P$ functions appearing in the four parametric equations above are integrals in a general null geodesic $\xi^{\alpha}$. Their definition is the same as in eq.~(\ref{n1}), but with $w_{|j}$ exchanged for $w$.

For $\delta = 0$, eq.~(\ref{k^v}) leads to a solution for $\alpha$ in terms of $\beta$:
\begin{equation}
\alpha = \frac{-\beta \epsilon_{\times}\, h_{\times} \pm i |\beta|}{1 + \epsilon_+\, h_+}\,,
\end{equation}
for which the only possible real solution is $\alpha= \beta = 0$. Replacing $\delta=\alpha=\beta=0$ in eqs.~(\ref{k^y}), (\ref{k^z}) and (\ref{u}) and integrating:
\begin{align}
\xi^y(\vartheta) = \xi^y(0)\,, \quad \xi^z(\vartheta) & = \xi^z(0)\,, \\
\xi^u(\vartheta) = \xi^u(0)\ \Rightarrow\ \xi^x(\vartheta) & = \xi^t(\vartheta) - t_{\mathcal{E}} + x_{\mathcal{S}}\,. \label{xdelta0}
\end{align}
Finally, noting that $\Gamma^t_{\mu \nu} = 0$ for $\mu,\nu = t, x$ and using the explicit geodesic equation for $\xi^t$:
\begin{equation}
\xi^t(\vartheta) = C \vartheta + \xi^t(0)\,,
\end{equation}
where $C$ is a constant. Once substituted in eq.~(\ref{xdelta0}):
\begin{equation}
\xi^x(\vartheta) = C\vartheta + \xi^x(0)\,.
\end{equation} 

We note that, in \cite{Rakhmanov2014}, the $\delta = 0$ solution is attributed only to non-null geodesics. Here we have shown its existence for null geodesics as well. It is the trajectory of a photon traveling purely along the GW propagation direction (here, the $x$ direction). We notice that this trajectory is not affected by the GW in any way: it is the same found in flat Minkowski spacetime. For this special case, the procedure used to calculate the radar distance in subsection \ref{subsec:unperturbed_spatial_trajectories} is in fact rigorous. A solution with these properties is expected, for example, by Lemma II found in p. 326 of \cite{Rindler2006}. 

Since there is no perturbation in the $\delta = 0$ case, the focus here will be on the $\delta \neq 0$ solutions. If our sole intention is to calculate the complete radar distance, there is no need to solve for $A$, $B$ and $\delta$ when imposing the mixed conditions to the parametric equations of the null geodesics, as exemplified by the treatment in \cite{Rakhmanov2009}. Here, however, the value of these constants will be explicitly computed so that the general parametric equations for rays 1 and 2 can be obtained and further aspects of the influence of GWs on light (for example, how the frequency evolves along the rays, in section \ref{sec:Doppler_effect}\,) can be thoroughly analyzed. 

The system arising from evaluating eqs.~(\ref{x}), (\ref{y}) and (\ref{z}) at $\vartheta_{\mathcal{R}}$, imposing the partial boundary conditions for ray 1, eqs.~(\ref{boundE}) and (\ref{boundR1}), and using eq.~(\ref{u}) to replace $ \vartheta_{\mathcal{R}} \delta$ for $\Delta x - \Delta t_{\mathcal{E, R}}$ results in:
\begin{align}
\Delta x &= \frac{1}{2} [(\Delta x - \Delta t_{\mathcal{E}, \mathcal{R}}) (1 - A^2_{|1} - B^2_{|1}) \,+  2\, \epsilon_{\times}\, m_{\times|1}(\vartheta_{\mathcal{R}}) A_{|1} B_{|1} \nonumber \\ & \hspace{145pt}+ \epsilon_+\, m_{+|1}(\vartheta_{\mathcal{R}})  (A^2_{|1} - B^2_{|1}) ]\,, \label{xbound} \\
\Delta y &= A_{|1} (\Delta x - \Delta t_{\mathcal{E}, \mathcal{R}}) - A_{|1} \epsilon_+\, m_{+|1}(\vartheta_{\mathcal{R}}) \,- B_{|1} \epsilon_{\times}\, m_{\times|1}(\vartheta_{\mathcal{R}})\,,  \label{ybound} \\
\Delta z &=  B_{|1} (\Delta x - \Delta t_{\mathcal{E}, \mathcal{R}}) + B_{|1} \epsilon_+\, m_{+|1}(\vartheta_{\mathcal{R}}) \,- A_{|1} \epsilon_{\times}\, m_{\times|1}(\vartheta_{\mathcal{R}})\,. \label{zbound}
\end{align}
Multiplying eq.~(\ref{ybound}) by $-A_{|1}/2$ and eq.~(\ref{zbound}) by $-B_{|1}/2$ and adding the resulting expressions yields:
\begin{align}
\epsilon_+ m_{+|1}(\vartheta_{\mathcal{R}}) \frac{(A^2_{|1} - B^2_{|1})}{2} = &-\frac{(A_{|1} \Delta y + B_{|1} \Delta z)}{2} + (\Delta x - \Delta t_{\mathcal{E}, \mathcal{R}})\frac{(A^2_{|1} + B^2_{|1})}{2}\nonumber \\
&  - \epsilon_{\times} m_{\times |1}(\vartheta_{\mathcal{R}})A_{|1} B_{|1} \,. 
\end{align}
Substituting for this into eq.~(\ref{xbound}) provides
\begin{equation}
\Delta t_{\mathcal{E, R}} = - (\Delta x + A_{|1} \Delta y + B_{|1} \Delta z)\,. \label{timeAB}
\end{equation}
Replacing this last expression in eqs.~(\ref{ybound}) and (\ref{zbound}), a system of two equations and two unknowns ($A_{|1}$ and $B_{|1}$) is attained. It will be solved perturbatively by writing 
\begin{align}
	A_{|1} = A_{(\bm{0})|1} + A_{+|1} \epsilon_+ + A_{\times|1} \epsilon_{\times}\,,
	\label{eq:A_perturbative_expansion}
\end{align}
and a similar expression for $B_{|1}$.  

In zero-th order ($\epsilon_+ = \epsilon_{\times} = 0$):
\begin{align}
&2A_{(\bm{0})|1} \Delta x + (A_{(\bm{0})|1}^2 - 1) \Delta y + A_{(\bm{0})|1} B_{(\bm{0})|1} \Delta z = 0\,, \label{A_0^2} \\
&2B_{(\bm{0})|1} \Delta x + A_{(\bm{0})|1} B_{(\bm{0})|1} \Delta y + (B_{(\bm{0})|1}^2 - 1) \Delta z = 0\,, \label{B_0^2}
\end{align}
whose solutions are:
\begin{equation}
A_{(\bm{0})|1} = \frac{\Delta y}{\Delta x - \Delta \ell} \;\; \text{and} \;\; B_{(\bm{0})|1} = \frac{\Delta z}{\Delta x - \Delta \ell}\,. 
\end{equation}

In first order, since $\epsilon_+$ and $\epsilon_{\times}$ are considered independent, we can solve for each of them separately. For plus polarization terms ($\epsilon_{\times} = 0$), replacing the values for $A_{(\bm{0})|1}$ and $B_{(\bm{0})|1}$, the system becomes 
\begin{equation}
	G \begin{bmatrix}
	A_{+|1} \\ B_{+|1}
	\end{bmatrix} = \begin{bmatrix}
	\Delta y \\ - \Delta z
	\end{bmatrix} m_{+|1} (\vartheta_{\mathcal{R}}),
\end{equation}
where
\begin{equation}
	G \leqdef \begin{bmatrix}
	(\Delta x - \Delta \ell)^2 + \Delta y^2 & \Delta y \Delta z \\ \Delta y \Delta z & (\Delta x - \Delta \ell)^2 + \Delta z^2
	\end{bmatrix}.
\end{equation}
Analogously, with only the cross polarization terms ($\epsilon_+=0$):
\begin{equation}
G \begin{bmatrix}
A_{\times|1} \\ B_{\times|1}
\end{bmatrix} = \begin{bmatrix}
\Delta z \\ \Delta y
\end{bmatrix} m_{\times|1} (\vartheta_{\mathcal{R}}).
\end{equation} 
Inverting these linear systems, the complete solutions for $A_{|1}$ and $B_{|1}$ are found to be:
\begin{align}
A_{|1} =\ & \frac{1}{\Delta x - \Delta \ell}\bigg\{\Delta y - \frac{1}{2\Delta \ell} \bigg[\Delta y \bigg(1 + \frac{2\Delta z^2}{(\Delta x - \Delta \ell)^2} \bigg) \epsilon_+\, m_{+|1}(\vartheta_{\mathcal{R}})  \nonumber \\ 
&\hspace{95pt} + \Delta z \bigg(1 + \frac{\Delta z^2 - \Delta y^2}{(\Delta x - \Delta \ell)^2} \bigg) \epsilon_{\times}\, m_{\times|1}(\vartheta_{\mathcal{R}}) \bigg]\bigg \}\,, \label{A1}\\
B_{|1}=\ & \frac{1}{\Delta x - \Delta \ell}\bigg\{\Delta z +  \frac{1}{2\Delta \ell} \bigg[\Delta z \bigg(1 + \frac{2\Delta y^2}{(\Delta x - \Delta \ell)^2} \bigg) \epsilon_+\, m_{+|1}(\vartheta_{\mathcal{R}})  \nonumber \\ 
&\qquad\qquad\! \hspace{55pt}-\Delta y \bigg(1 + \frac{\Delta y^2 - \Delta z^2}{(\Delta x - \Delta \ell)^2} \bigg) \epsilon_{\times}\, m_{\times|1}(\vartheta_{\mathcal{R}}) \bigg]\bigg \}\,. \label{B1}
\end{align}
Replacing them on eq.~(\ref{timeAB}) and comparing to eq.~(\ref{tR-tE}):
\begin{equation}
	\Delta t_{\mathcal{E, R}} = \Delta t_{\mathcal{E}, \tilde{\mathcal{R}}}. \label{sametime}
\end{equation}
For ray 2, the calculation is similar, being only necessary to make the changes $\Delta x^i \rightarrow -\Delta x^i$ and   $m_{P|1}(\vartheta_{\mathcal{R}}) \rightarrow m_{P|2}(\vartheta_{\mathcal{D}})$, resulting in 
\begin{equation}
	\Delta t_{\mathcal{R, D}} = \Delta t_{{\tilde{\mathcal{R}}}, \tilde{\mathcal{D}}}.
\end{equation}
As a consequence
\begin{equation}
	D_{R,\xi} = D_{R,\tilde{\xi}}. 
\end{equation}

This may sound as an unexpected coincidence, since the null geodesics in Minkowski spacetime and those in a GW one with $\delta \neq 0$ are not the same family of curves. In \cite{Rakhmanov2009}, this coincidence is noticed without the explicit calculation of the constants of motion in terms of the known parameters $\bm{P}$. However, no discussion was made on which hypotheses are behind such a result. In \cite{Finn2009}, these hypotheses are discussed in a more general context, but we will elaborate on them for a particular circumstance. Is this coincidence in radar distance a consequence of some GW property, or is it more general? This will be the topic of subsection \ref{subsec:paramet_equations}. 

We can rewrite $D_{R,\xi}$ by introducing angles $\theta$ and $\phi$ such that: 
\begin{align}
\Delta x = \Delta \ell \cos{\theta}\,, \quad\Delta y = \Delta \ell \sin{\theta} \cos{\phi}\,, \quad \Delta z = \Delta \ell \sin{\theta} \sin{\phi}. \label{sph} 
\end{align}

\begin{figure}[t]
	\centering
	\includegraphics[scale = 0.15]{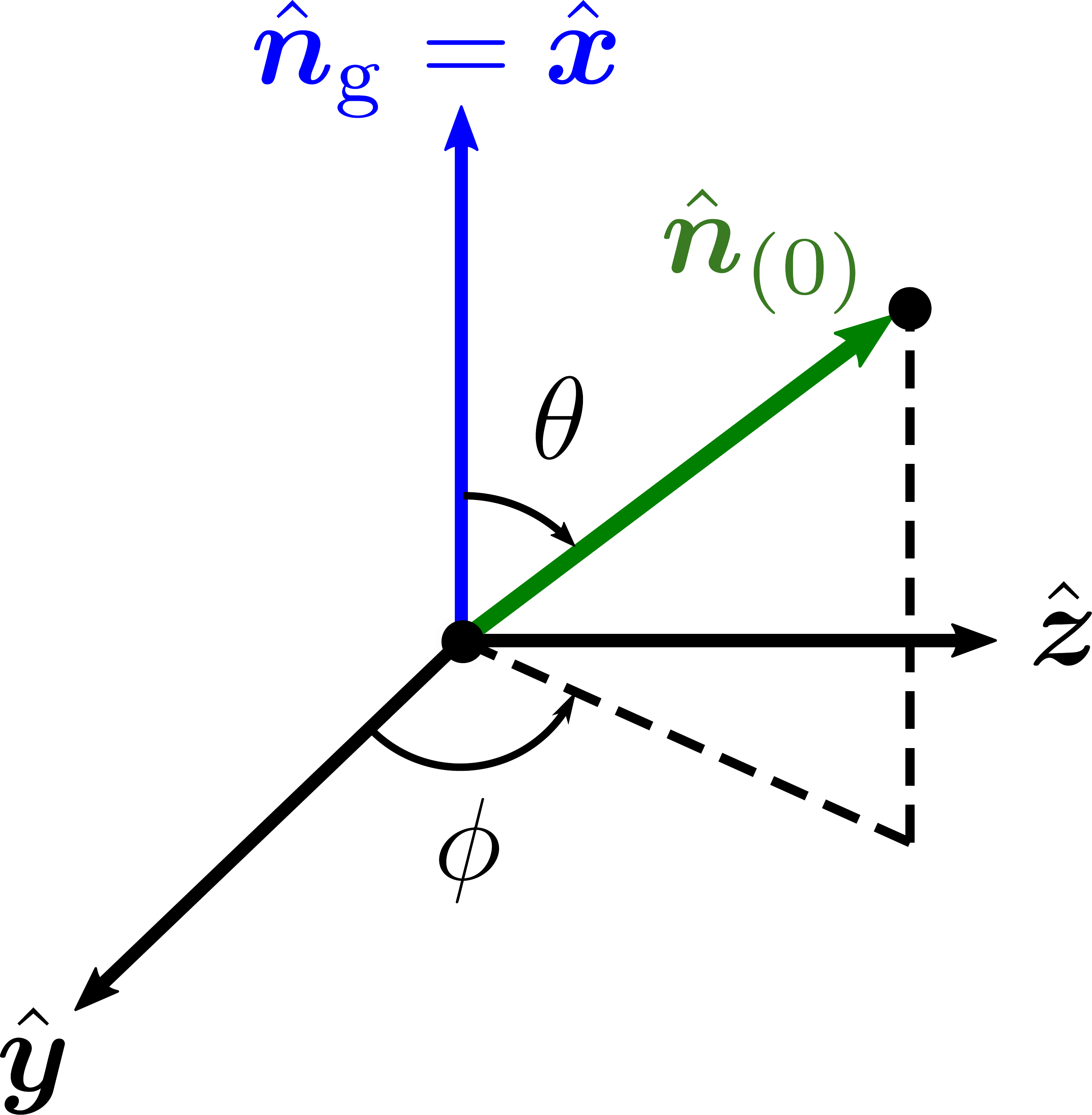}
	\caption{Three-dimensional rest space of an instantaneous observer immersed in the GW region. The spatial direction of propagation of the gravitational wave, {$\hat{\boldsymbol n}_{\textrm{g}}$}, is chosen along the local unit vector $\hat{{\boldsymbol x}}$, whereas the local zero-th order spatial direction of the laser beam is $\hat{\boldsymbol n}_{(\bm{0})}$\,.} \label{fig:arm_direction}
\end{figure}
Replacing this in eq.~(\ref{D_R}) gives:
\begin{align}
D_{R, \xi}(t_{\mathcal{Q}}) =\, \Delta \ell 
& -\frac{\epsilon_+}{2} \left[ \cos^2\left(\frac{\theta}{2}\right)M_{+|1}(t_{\mathcal Q}) +  \sin^2\left( \frac{\theta}{2}\right) M_{+|2}(t_{\mathcal Q}) \right]\cos 2\phi {} \nonumber \\
&  -\frac{\epsilon_\times}{2} \left[ \cos^2\left( \frac{\theta}{2} \right)M_{\times|1}(t_{\mathcal Q}) + \sin^2\left( \frac{\theta}{2} \right)M_{\times|2}(t_{\mathcal Q}) \right]\sin 2\phi \,. \label{radardistspheric}
\end{align}
If observers $\mathcal{S}$ and $\mathcal{M}$ are the extremities of an interferometer arm, the angle $\theta$ may be identified with the angle between the GW propagation direction and the zero-th order arm direction, as depicted in figure~\ref{fig:arm_direction}.
  
\subsection{The coincidence}
\label{subsec:paramet_equations}
  
The content of this subsection is in consonance with section IV of \cite{Finn2009}. The author considers generic perturbations over an arbitrary background metric and lists a set of conclusions, although not explicitly derived, regarding the necessary hypotheses behind $D_{R, \xi} = D_{R,\tilde{\xi}}$. Here we make a more explicit discussion, for a slightly more general metric than in eq.~(\ref{line_elem_TT}), namely
\begin{equation}
g_{\alpha \beta}(x^{\mu}, \epsilon) =  \eta_{\alpha \beta} + \epsilon h_{ \alpha \beta}(x^{\mu})\,,\quad h_{\alpha t}=0\,, \label{weak_field}
\end{equation}
but still work with the comoving observers (\ref{TTframe}) and with the same mixed conditions (\ref{boundE} -- \ref{continuity}).

We begin by defining for any quantity $V$ the variation 
\begin{equation}
	\epsilon \Delta_{\epsilon} V \leqdef V - V_{(0)}\,.
\end{equation}
Then, rewriting eq.~(\ref{timeERunp}) but now for the perturbed spatial trajectories, replacing 
\begin{align} 
k^i_{|1}&= k^i_{(0)|1} + \epsilon \Delta_{\epsilon} k^i_{|1}\,, \\ \vartheta_{\mathcal{R}} &= \vartheta_{\mathcal{R}(0)} + \epsilon \Delta_{\epsilon} \vartheta_{\mathcal{R}}\,,
\end{align}
and expanding the resulting expression, we obtain:
\begin{align}
	\Delta t_{\mathcal{E}, \mathcal{R}} &= \int^{\vartheta_{\mathcal{R}}}_{0} \sqrt{[\delta_{ij} + \epsilon h_{ij}(w_{|1}(\vartheta))] k^i_{|1} k^j_{|1}} d\vartheta \nonumber \\& = \Delta t_{\mathcal{E}, \tilde{\mathcal{R}}} + \epsilon \sqrt{\delta_{ij}k^i_{(0)|1} k^j_{(0)|1}}  \Delta_{\epsilon} \vartheta_{\mathcal{R}} +\frac{\delta_{ij} k^i_{(0)|1}}{\sqrt{\delta_{ml}k^m_{(0)|1} k^l_{(0)|1}}}  \int_{0}^{\vartheta_{\mathcal{R}_{(0)}}} \epsilon \Delta_{\epsilon} k^j d\vartheta, \label{Deltaeps}
\end{align} 
remembering that $\vartheta_{\mathcal{R}(0)}=\vartheta_{\mathcal{R}_{(0)}} = \vartheta_{\tilde{\mathcal{R}}}$. The fact that, in general, $\Delta_{\epsilon} \vartheta_{\mathcal{R}} \neq 0$ is further explored in appendix \ref{app:domains}. Moreover, by definition:
\begin{align}
\epsilon \Delta_{\epsilon} \bigg(\int_{0}^{\vartheta_{\mathcal{R}}} k^j_{|1} d\vartheta \bigg) &\leqdef \int_{0}^{\vartheta_{\mathcal{R}}} k^j_{|1} d\vartheta - \int_{0}^{\vartheta_{\mathcal{R}_{(\bm{0})}}} k^j_{(0)|1} d\vartheta   \nonumber \\ &= k^j_{(0)|1} \epsilon \Delta_{\epsilon} \vartheta_{\mathcal{R}} + \int_{0}^{\vartheta_{\mathcal{R}_{(0)}}} \epsilon \Delta_{\epsilon} k^j_{|1} d\vartheta.
\end{align}
Replacing this in the last term of eq.~(\ref{Deltaeps}):
\begin{align}
	\Delta t_{\mathcal{E}, \mathcal{R}} &= \Delta t_{\mathcal{E}, \tilde{\mathcal{R}}} + \frac{\delta_{ij} k^i_{(0)|1}}{\sqrt{\delta_{ml}k^m_{(0)|1} k^l_{(0)|1}}} \epsilon \Delta_{\epsilon} \bigg(\int_{0}^{\vartheta_{\mathcal{R}}} k^j_{|1} d\vartheta \bigg) \nonumber \\ &=  \Delta t_{\mathcal{E}, \tilde{\mathcal{R}}} + \delta_{ij} \frac{\Delta x^i}{\Delta \ell} \epsilon \Delta_{\epsilon}  (\Delta x^j). \label{tERdiff}
\end{align}
where on the last equality, the first of eqs.~(\ref{k_{(0)}}) was used.

Likewise, for the back-trip:
\begin{align}
	\Delta t_{\mathcal{R}, \mathcal{D}} &= \Delta t_{\tilde{\mathcal{R}}, \tilde{\mathcal{D}}} - \frac{\delta_{ij} k^i_{(0)|2}}{\sqrt{\delta_{ml}k^m_{(0)|2} k^l_{(0)|2}}} \epsilon \Delta_{\epsilon} (\Delta x^j) \nonumber \\ & = \Delta t_{\tilde{\mathcal{R}}, \tilde{\mathcal{D}}} + \delta_{ij} \frac{\Delta x^i}{\Delta \ell} \epsilon \Delta_{\epsilon} (\Delta x^j). \label{tRDdiff}
\end{align}
Note that, since eqs.~(\ref{tERdiff}) and (\ref{tRDdiff}) are valid for any weak gravitational field with $h_{\alpha t} = 0$, it is not necessary to impose any GW property or choose a specific gauge. The quantities expressed in these equations only are directly related to the radar distance because $h_{tt} = 0$, since, without that assumption, the coordinate difference $t_{\mathcal{D}} - t_{\mathcal{E}}$ would not be trivially equal to the difference in proper time measured by $\mathcal{S}$. Summing eqs.~(\ref{tERdiff}) and (\ref{tRDdiff}) leads to:
\begin{align}
	D_{R, \xi} - D_{R, \tilde{\xi}} = 2\epsilon \delta_{ij} \frac{\Delta x^i}{\Delta \ell_{(0)}}  \Delta_{\epsilon}(\Delta x^j) = 2 \epsilon \Delta_{\epsilon}(\Delta \ell),
\end{align}
From such a difference, notice that the coincidence happens because the fixed spatial coordinates of each chosen observer are not $\epsilon$-dependent and so
\begin{equation}
	\Delta_{\epsilon}(\Delta x^i) = 0\,. \label{Deltaepsilon0}
\end{equation}
But this is, contrary to what is concluded in \cite{Finn2009}, only a sufficient condition for the coincidence to occur. The necessary condition is that the coordinate distance traveled by light $\Delta \ell$ must be independent of $\epsilon$.

For the particular GW case, with observers adapted to the TT gauge, (\ref{Deltaepsilon0}) is a direct consequence of the well known result that observers with $x^i(\vartheta) = \text{const}$ in the absence of GWs (when $\epsilon = 0$) remain with these same coordinate values when GWs are present ($\epsilon \neq 0$) \cite{Rindler2006, Maggiore2007}.  

Although $\xi_{|j} \neq \tilde{\xi}_{|j}$, if $\Delta_{\epsilon}(\Delta x^i) = 0$, we can conclude that events $\tilde{\mathcal{R}}$ and $\tilde{\mathcal{D}}$ would, respectively, have the same time coordinate of events $\mathcal{R}$ and $\mathcal{D}$, because of eqs.~(\ref{tERdiff}), (\ref{tRDdiff}). Then, since the partial boundary conditions (\ref{boundR1}) and (\ref{boundD}) fix the same spatial coordinates for them as well, we conclude that, pictorially, observing figure~\ref{fig:radar_distance_3_types_of_curve}, these events would coincide.
\section{Frequency shift}
\label{sec:Doppler_effect}

The frequency shift of light in a GW spacetime was originally discussed in \cite{Kaufmann1970}. It is usually studied in the context of Doppler tracking of spacecrafts \cite{Estabrook1975, Tinto1998, Armstrong2006} and when discussing laser frequency noise in the LISA interferometer \cite{Tinto2002}, for example. Here, our ultimate aim is to study this effect in a way that its contributions to the fluctuations in the final intensity pattern of an interferometric process, calculated in L2, can be properly identified. 

From now on, only the perturbed, red, solid curves of figure \ref{fig:radar_distance_3_types_of_curve} will be used. Starting from eq.~(\ref{initfreq}), we will derive the frequency shift after a complete round-trip of light between $\mathcal{S}$ and $\mathcal{M}$. To achieve this goal, the constants $\delta_{|1}$ and $\delta_{|2}$ will be determined, allowing one to obtain the general parametric expressions for the null rays (subsection \ref{subsec: gen_parametric_eq}).

In subsection \ref{subsec:radar_doppler_effect} we predict the frequency shift for a non-monochromatic gravitational plane wave and interpret its origin in terms of the radar distance concept previously presented as the usual Doppler effect. 

In addition, in subsection \ref{subsec:freq_vs_radar_dist} we address one of the most frequent questions regarding the detection of gravitational waves through interferometry: ``If a GW stretches the arm and the laser wavelength simultaneously, should not these effects cancel each other? How can we detect GWs then?". In \cite{Saulson1997}, a qualitative discussion is made without the use of an explicit mathematical formalism. In \cite{Faraoni2007}, a quantitative discussion is made, but only in the long GW wavelength limit. In this work, we study the question in a more general sense, relying only on the electromagnetic geometrical optics laws as the mathematical framework for light propagation.

\subsection{The general parametric equations for ray 1} \label{subsec: gen_parametric_eq}

As part of the mixed conditions assumed in our analysis, the initial frequency measured by the source $\mathcal{S}$ is given by eq.~(\ref{initfreq}), which, differentiating eq.~(\ref{t}) with respect to $\vartheta$ and evaluating in $\mathcal{E}$ becomes:
\begin{align}
	&{\omega_{\textrm{e}}}_{\mathcal{E}} = - \frac{\delta_{|1}}{2} [1 + A^2_{|1} + B^2_{|1} + 2\epsilon_{\times}{h_{\times|1}}(0)A_{|1}B_{|1} + \epsilon_+(A^2_{|1} - B^2_{|1}) h_{+|1}(0)],  \label{init_freq}
\end{align}
where $h_{P|j}(\vartheta) \leqdef h_P(\xi^t_{|j}(\vartheta) - \xi^x_{|j}(\vartheta)) $, and eq.~(\ref{u}) was used to calculate the derivatives of $m_{P|j}$:
\begin{equation}
	\frac{dm_{P|j}}{d \vartheta} = - h_{P|j}  \delta_{|j}. \label{np_deriv}
\end{equation}
Expression (\ref{init_freq}) can be inverted to give the value of the constant $\delta_{|1}$. Replacing eqs.~(\ref{A1}) and (\ref{B1}) subsequently:
\begin{align}
	&\delta_{|1} = - {\omega_{\textrm{e}}}_{\mathcal{E}} \bigg\{1 - \frac{\Delta x}{\Delta\ell} + \frac{1}{\Delta\ell^3} \bigg[\epsilon_+ \left(m_{+|1}(\vartheta_{\mathcal{R}}) - \Delta\ell\, h_{+|1}(0)\right) \frac{(\Delta y^2 - \Delta z^2)}{2}  \nonumber \\ & \hspace{150pt} +\epsilon_{\times} \left(m_{\times|1}(\vartheta_{\mathcal{R}}) - \Delta\ell\, h_{\times|1}(0)\right) \Delta y \Delta z  \bigg]  \bigg\}. \label{delta}
\end{align}

Particularizing eqs.~(\ref{t}) and (\ref{x}) to ray 1 and replacing on them eqs.~(\ref{A1}), (\ref{B1}) and (\ref{delta}), the general parametric equations for ray 1 are found:
\begin{align}
\xi^x_{|1}(\vartheta) =  {x_{}}_{\mathcal{S}} & +  \frac{\vartheta {\omega_{\textrm{e}}}_{\mathcal{E}}}{\Delta\ell}\bigg\{\Delta x - \epsilon_+ \bigg[\frac{h_{+|1}(0) \Delta x}{\Delta\ell - \Delta x} + \frac{m_{+|1}(\vartheta_{\mathcal{R}})}{\Delta\ell}\bigg] \frac{(\Delta y^2 - \Delta z^2)}{2\Delta\ell}  \nonumber \\ &\hspace{90pt}-\epsilon_{\times} \bigg[\frac{ h_{\times|1}(0) \Delta x}{\Delta\ell - \Delta x} + \frac{m_{\times|1}(\vartheta_{\mathcal{R}})}{\Delta\ell}\bigg] \frac{\Delta y \Delta z}{\Delta\ell} \bigg\}  \nonumber \\
& +\frac{1}{(\Delta\ell- \Delta x)^2} \bigg[\epsilon_+ m_{+|1}(\vartheta) \frac{\Delta y^2 - \Delta z^2}{2} + \epsilon_{\times} m_{\times|1} (\vartheta) \Delta y \Delta z  \bigg]\,,   \label{x_ray_1} \\ \nonumber \\
\xi^y_{|1}(\vartheta) = {y_{}}_{\mathcal{S}} & + \frac{\vartheta {\omega_{\textrm{e}}}_{\mathcal{E}}}{\Delta\ell}\bigg\{\Delta y  + \frac{\Delta y}{\Delta\ell(\Delta x - \Delta\ell)}   \left[\epsilon_{\times} \Delta y \Delta z  \, h_{\times|1}(0) + \epsilon_+ \frac{(\Delta y^2 - \Delta z^2)}{2}  h_{+|1}(0)\right] \nonumber \\ & \hspace{62pt}+ \frac{\epsilon_{\times}m_{\times|1}(\vartheta_{\mathcal{R}})\Delta z}{\Delta\ell^2(\Delta x - \Delta\ell)^2} \left[\Delta y^2 \Delta x  + (\Delta x-\Delta \ell)(\Delta\ell^2-2\Delta y ^2)\right]   \nonumber \\ & \hspace{62pt} + \frac{\epsilon_+ m_{+|1}(\vartheta_{\mathcal{R}}) \Delta y}{2\Delta\ell^2(\Delta x - \Delta\ell)^2}\left[ \Delta\ell^2 \Delta x + (\Delta x - 2\Delta\ell)  (\Delta x^2 + 2 \Delta z^2)\right] \hspace{-4pt}\bigg\} \nonumber \\ & +\frac{1}{\Delta\ell - \Delta x} \left[\epsilon_+ m_{+|1}(\vartheta) \Delta y + \epsilon_{\times} m_{\times|1}(\vartheta) \Delta z \right] ,\label{y_ray_1} \\ \nonumber 
\end{align}
\begin{align}
\xi^z_{|1}(\vartheta) = {z_{}}_{\mathcal{S}} & +  \frac{\vartheta {\omega_{\textrm{e}}}_{\mathcal{E}}}{\Delta\ell}\bigg\{\Delta z + \frac{\Delta z}{\Delta\ell(\Delta x - \Delta\ell)} \left[\epsilon_{\times} \Delta y \Delta z  \, h_{\times|1}(0) 
+ \epsilon_+ \frac{(\Delta y^2 - \Delta z^2)}{2}  h_{+|1}(0) \right]  \nonumber \\ & \hspace{60pt} - \frac{\epsilon_{\times}m_{\times|1}(\vartheta_{\mathcal{R}})\Delta y}{\Delta\ell^2(\Delta x - \Delta\ell)^2}\left[\Delta z^2 \Delta x  + (\Delta x-\Delta \ell)(\Delta\ell^2-2\Delta z^2)\right]  \nonumber\\& \hspace{60pt}  + \frac{\epsilon_+ m_{+|1} (\vartheta_{\mathcal{R}}) \Delta z}{2\Delta\ell^2(\Delta x - \Delta\ell)^2}\left[ \Delta\ell^2 \Delta x + (\Delta x - 2\Delta\ell)  (\Delta x^2 + 2 \Delta y^2)\right] \hspace{-4pt}\bigg\} \nonumber\\& +\frac{1}{\Delta\ell - \Delta x}  \left[\epsilon_{\times} m_{\times|1}(\vartheta) \Delta y - \epsilon_+ m_{+|1}(\vartheta) \Delta z\right] ,\label{z_ray_1} \\ \nonumber \\
\xi^t_{|1}(\vartheta) = t_{\mathcal{E}} & + \frac{\vartheta {\omega_{\textrm{e}}}_{\mathcal{E}}}{\Delta \ell} \bigg\{\Delta \ell - \frac{1}{(\Delta\ell - \Delta x)} \bigg[\epsilon_{\times} \Delta y \Delta z h_{\times|1}(0) + \epsilon_+ \frac{(\Delta y^2 - \Delta z^2)}{2} h_{+|1}(0)  \bigg]\bigg\} \nonumber \\ & + \frac{1}{(\Delta\ell - \Delta x)^2}  \bigg[\epsilon_{\times} m_{\times|1}(\vartheta) \Delta y \Delta z  + \epsilon_+ m_{+|1}(\vartheta) \frac{(\Delta y^2 - \Delta z^2)}{2}\bigg]  .\label{t_ray_1} 
\end{align}
The equations for ray 2 are obtained by exchanging ${x{}}^i_{\mathcal{S}} $ and $x^i_{\mathcal{M}}$, so that $\Delta x^i \rightarrow -\Delta x^i$; additionally, we also change $m_{P|1}(\vartheta_{\mathcal{R}}) \rightarrow m_{P|2}(\vartheta_{\mathcal{D}})$, $h_{P|1}(0) \rightarrow h_{P|2}(0)$ and ${\omega_{\textrm{e}}}_{\mathcal{E}}\rightarrow \omega_{\textrm{e}|2}(0) = \omega_{\textrm{e}|1}(\vartheta_{\mathcal{R}})$ on the above equations. 

With the value of $\delta_{|1}$, $\vartheta_{\mathcal{R}}$ can also be determined. Evaluating (\ref{u}) at $\vartheta_{\mathcal{R}}$, using eqs.~(\ref{tR-tE}) and (\ref{sametime}), together with (\ref{delta}):
\begin{align}
	&\vartheta_{\mathcal{R}} = \frac{\Delta x - \Delta t_{\mathcal{E,R}}}{\delta_{|1}} = \frac{1}{\omega_{e\mathcal{E}}} \bigg\{ \Delta \ell + \frac{1}{(\Delta \ell - \Delta x)} \bigg[\epsilon_{\times} K_{\times} \Delta y \Delta z + \epsilon_+ K_+ \frac{\Delta y^2 -  \Delta z^2}{2}\hspace{-1pt} \bigg]   \hspace{-3pt}\bigg\}.
\end{align}
where, remembering ($\ref{M_P_1}$) and ($\ref{M_P_2}$), 
\begin{align}
	K_P \leqdef \,& \frac{\Delta x - 2 \Delta \ell}{\Delta \ell(\Delta \ell - \Delta x)}m_{P|1}(\vartheta_{\mathcal{R}(\bm{0})}) + h_{P|1}(0) \nonumber \\ = \,& \frac{\Delta x - 2 \Delta \ell}{\Delta \ell(\Delta \ell - \Delta x)}M_{P|1}(t_{\mathcal{D}} - \Delta \ell) +  h_{P}(t_{\mathcal{D}} - 2\Delta \ell - x_{\mathcal{S}}).
\end{align}
Making the already mentioned changes to calculate the back-trip analogue of $\vartheta_{\mathcal{R}}$ (the frequency $\omega_{\textrm{e}|1}(\vartheta_{\mathcal{R}})$ will be calculated in eq.~($\ref{omega1}$)), we find
\begin{align}
	&\vartheta_{\mathcal{D}} = \frac{1}{\omega_{e\mathcal{E}}} \bigg[ \Delta \ell +\bigg(\epsilon_{\times} Q_{\times} \Delta y \Delta z + \epsilon_+ Q_+ \frac{\Delta y^2 -  \Delta z^2}{2}\hspace{-1pt} \bigg)   \hspace{-3pt}\bigg],
\end{align}
where
\begin{align}
	Q_P \leqdef \, & \frac{h_{P|2}({\vartheta_{\mathcal{R}}}_{(\bm{0})})}{\Delta \ell + \Delta x} - \frac{\Delta h_{P|1}}{\Delta \ell - \Delta x} - \frac{\Delta x + 2 \Delta \ell}{\Delta \ell(\Delta \ell + \Delta x)^2}m_P({\vartheta_{\mathcal{D}}}_{(\bm{0})})
	\nonumber \\ = \,& \frac{h_{P}(t_{\mathcal{D}} - \Delta \ell - x_{\mathcal{M}})}{\Delta \ell + \Delta x} - \frac{\Delta h_{P|1}}{\Delta \ell - \Delta x} - \frac{\Delta x + 2 \Delta \ell}{\Delta \ell(\Delta \ell + \Delta x)^2}M_{P|2}(t_{\mathcal{D}} - \Delta \ell)
\end{align}
with $\Delta h_{P|1} \leqdef h_{P|1}(\vartheta_{\mathcal{R}}) - h_{P|1}(0)$.
These expressions will be useful in L2. A further discussion about the domains of the parametrized curves and their split in unperturbed and perturbed parts is held in appendix \ref{app:domains}.

\subsection{Doppler effect} \label{subsec:radar_doppler_effect}

Differentiating eq.~(\ref{t_ray_1}) with respect to $\vartheta$, with the help of eqs.~(\ref{np_deriv}) and (\ref{delta}), the frequency of light along ray 1 is obtained. Evaluating it at event $\mathcal{R}$ results in:
\begin{align}
&\omega_{\textrm{e}|1}(\vartheta_{\mathcal{R}}) =  {\omega_{\textrm{e}}}_{\mathcal{E}} \bigg\{1 + \frac{1}{\Delta\ell(\Delta\ell - \Delta x)}  \bigg[\epsilon_+ [h_{+|1}(\vartheta_{\mathcal{R}}) - h_{+|1}(0)]  \frac{(\Delta y^2 - \Delta z^2)}{2}  \nonumber \\ &\hspace{180pt}+ \epsilon_{\times} [h_{\times|1}(\vartheta_{\mathcal{R}})  - h_{\times|1}(0)]   \Delta y \Delta z \bigg]\bigg\}. \label{omega1}	
\end{align}
For ray 2, given the continuity of the frequency at the reflection event, eq.~(\ref{continuity}), a similar expression is found:
\begin{align}
	\omega_{\textrm{e}|2}(\vartheta_{\mathcal{D}}) & =  \omega_{\textrm{e}|2} (0) \bigg\{1 + \frac{1}{\Delta\ell(\Delta\ell + \Delta x)} \bigg[\epsilon_+ [h_{+|2}(\vartheta_{\mathcal{D}}) - h_{+|2}(0)] \frac{(\Delta y^2 - \Delta z^2)}{2} \nonumber \\ & \hspace{220pt}+ \epsilon_{\times} [h_{\times|2}(\vartheta_{\mathcal{D}})  - h_{\times|2}(0)] \Delta y \Delta z  \bigg]\bigg\} \nonumber \\ 
	& = {\omega_{\textrm{e}}}_{\mathcal{E}} \bigg\{1 + \frac{1}{\Delta\ell} \bigg[\epsilon_+ \frac{(\Delta y^2 - \Delta z^2)}{2} \bigg( \frac{h_{+|2}(\vartheta_{\mathcal{D}})- h_{+|2}(0)}{\Delta\ell + \Delta x} +  \frac{h_{+|1}(\vartheta_{\mathcal{R}})- h_{+|1}(0)}{\Delta\ell - \Delta x} \bigg) \nonumber \\ 
	&\hspace{102pt} + \epsilon_{\times}\Delta y \Delta z \bigg( \frac{h_{\times|2}(\vartheta_{\mathcal{D}})- h_{\times|2}(0)}{\Delta\ell + \Delta x} +  \frac{h_{\times|1}(\vartheta_{\mathcal{R}})- h_{\times|1}(0)}{\Delta\ell - \Delta x}\bigg) \bigg] \bigg\}.\label{omega2}
\end{align}  
Then, in terms of the spherical coordinates introduced in eq.~(\ref{sph}),  the frequency shift measured by observer $\mathcal{S}$ after a round-trip of light is:
\begin{align}
	\frac{\Delta \omega_{\textrm{e}}}{{\omega_{\textrm{e}}}_{\mathcal{E}}} (t_{\mathcal{D}}) \leqdef \, & \frac{\omega_{\textrm{e}}(\vartheta_{\mathcal{D}}) - {\omega_{\textrm{e}}}_{\mathcal{E}}}{{\omega_{\textrm{e}}}_{\mathcal{E}}}\nonumber \\ = \, & \big \{\epsilon_+ [h_+(t_{\mathcal{D}} - x_{\mathcal{S}}) - h_+(t_{\mathcal{D}} - \Delta\ell - x_{\mathcal{M}})] \cos{2\phi} \, \nonumber \\ & + \epsilon_{\times}[h_{\times}(t_{\mathcal{D}} - x_{\mathcal{S}}) - h_{\times}(t_{\mathcal{D}} - \Delta\ell - x_{\mathcal{M}})] \sin{2\phi} \big\} \sin^2{\Big(\frac{\theta}{2}\Big)} \nonumber \\  & + \big\{\epsilon_+[h_+(t_{\mathcal{D}} - \Delta\ell - x_{\mathcal{M}})  -h_+(t_{\mathcal{D}} - 2\Delta\ell - x_{\mathcal{S}})]  \cos{2\phi}\, \nonumber \\ &  + \epsilon_{\times}[ h_{\times}(t_{\mathcal{D}} - \Delta\ell - x_{\mathcal{M}}) -h_{\times}(t_{\mathcal{D}} - 2\Delta\ell - x_{\mathcal{S}})] \sin{2\phi}\big\} \cos^2{\Big(\frac{\theta}{2}\Big)}  . \label{redshift}
\end{align}

This expression is in agreement with the ones presented in works like \cite{Armstrong2006, Estabrook1975}. It is the percentage difference either between initial and final frequency of light in a round-trip, or between the perturbed and unperturbed frequency at the final event.

From the purely (infinitesimal) shearing kinematics of the TT frame induced by the GWs (cf. L2), to which observers $\mathcal{S}$ and $\mathcal{M}$ belong, that manifests itself by the radar distance change with time, there is a clear indication that the above Doppler effect should exist. As pointed out in e.g. \cite{Ellis1971}, the frequency evolution along a ray relates to the kinematics of a given frame by
\begin{equation}
	\frac{1}{\omega_e}\frac{d \omega_e}{d \vartheta} = - \left(\frac{1}{3} \Theta + a_{\mu}n^{\mu}+\sigma_{\mu\nu}n^{\mu}n^{\nu}\right)\omega_e
\end{equation}
where $\Theta$ is the expansion scalar, $a^{\mu}$ is the four acceleration, $\sigma_{\mu \nu}$ is the shear tensor and $-n^{\mu}$ is the photon line of sight relative to the reference frame $u^{\mu}$ who measures the frequency. In fact, the change in radar distance and the frequency shift are actually two facets of the same GW-photon interaction. Note first that:
\begin{align}
&\epsilon_P \frac{d M_{P|1}}{dt_{\mathcal{Q}}}(t_{\mathcal{Q}}) = \epsilon_P[h_P(t_{\mathcal{D}} - \Delta\ell - x_{\mathcal{M}}) - h_P(t_{\mathcal{D}} - 2\Delta\ell - x_{\mathcal{S}})]
\end{align}   
and, similarly,
\begin{align}
\epsilon_P \frac{d M_{P|2}}{dt_{\mathcal{Q}}}(t_{\mathcal{Q}}) = \epsilon_P[h_P(t_{\mathcal{D}} + \Delta\ell - x_{\mathcal{S}}) - h_P(t_{\mathcal{Q}} - x_{\mathcal{M}})].
\end{align} 
Then, by differentiating eq.~(\ref{D_R}) with respect to $t_{\mathcal{Q}}$, one concludes that
\begin{equation}	
\frac{d D_{R,\xi}}{d t_{\mathcal{Q}}} (t_{\mathcal{Q}}(t_{\mathcal{D}})) = - \frac{1}{2} \frac{\Delta \omega_{\textrm{e}}}{\omega_{\textrm{e}\mathcal{E}}}(t_{\mathcal{D}}). \label{Doppler}
\end{equation}
This is indeed the usual expression describing a Doppler effect when relative velocities are small compared to the speed of light in vacuum (here $c = 1$), if we interpret the covariant radar velocity defined by
\begin{equation}
v_{R} \leqdef \frac{d D_{R,\xi}}{d\tau_{\mathcal{Q}}}  = \frac{d D_{R,\xi}}{dt_{\mathcal{Q}}}  \label{radar_velocity}
\end{equation}
as a relative velocity of observer $\mathcal{M}$ with respect to $\mathcal{S}$. Relation ($\ref{Doppler}$) can be trivially obtained in any spacetime by taking the ratio between the proper time differences of two emitted wave crests and the subsequently received ones \cite{Schutz2009}; the factor 1/2 arises from the accumulation of Doppler effects in the two rays in the round-trip.

In a general context, where the distance between two observers is not small compared to the GW wavelength, their relative motion cannot be described in terms of geodesic deviations and, moreover, in this non-local framework, a relative velocity cannot be uniquely defined by means of a comparison between velocities at different events, given the curved character of spacetime. On the other hand, eqs.~(\ref{Doppler}) and (\ref{radar_velocity}) highlight how suitable is the description of the relative motion between observers $\mathcal{M}$ and $\mathcal{S}$ in terms of the radar distance, regardless of how far they are from each other, in a covariant manner. Although $v_{R}$ plays the role of a relative velocity in the above expression, it was not necessary to compare the 4-velocities of the possibly distant observers, since it is measured solely in terms of information obtained by $\mathcal{S}$.

\begin{figure}[t] \label{fig:binary_waveform}
	\includegraphics[scale=0.5]{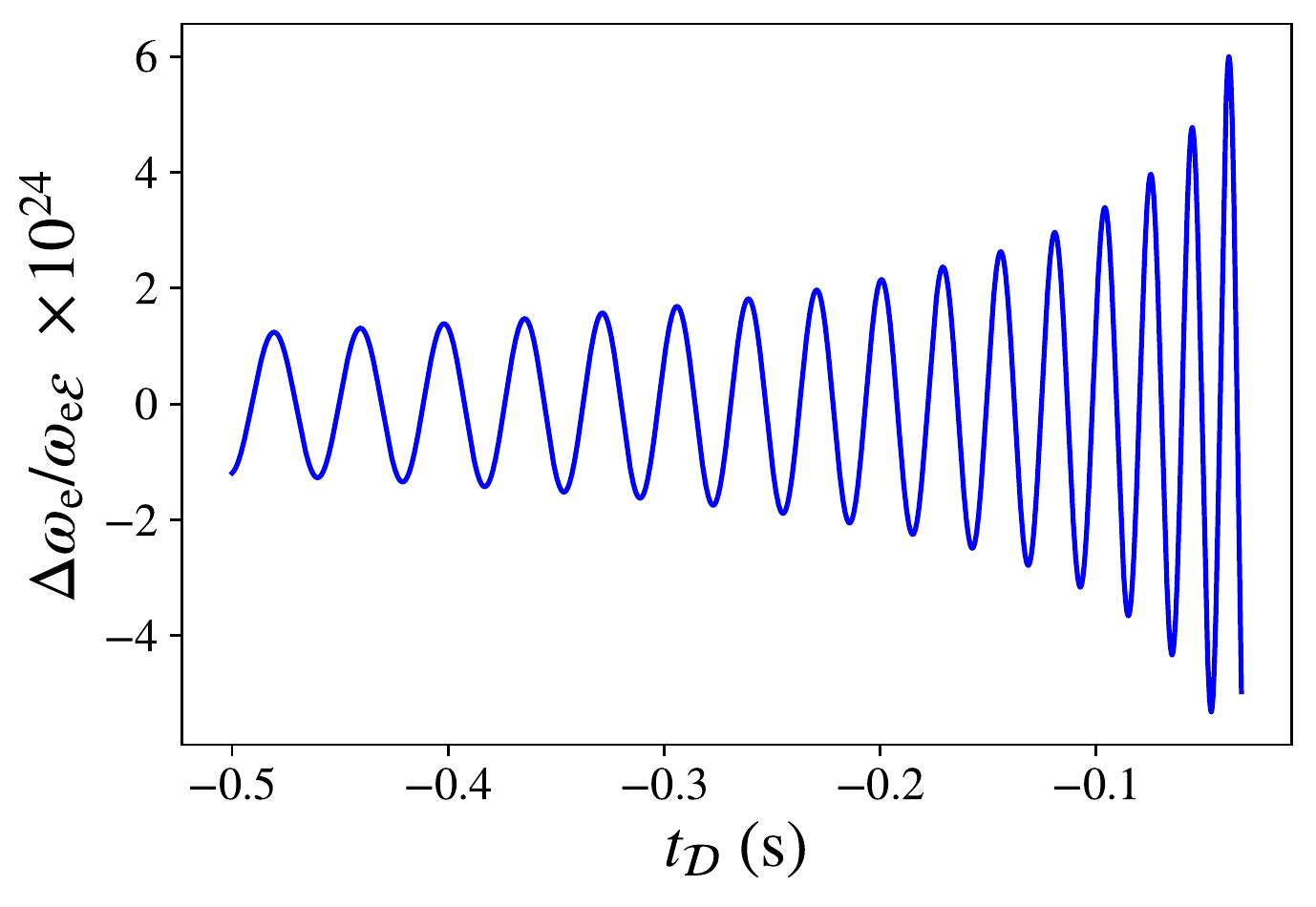}
	\includegraphics[scale=0.5]{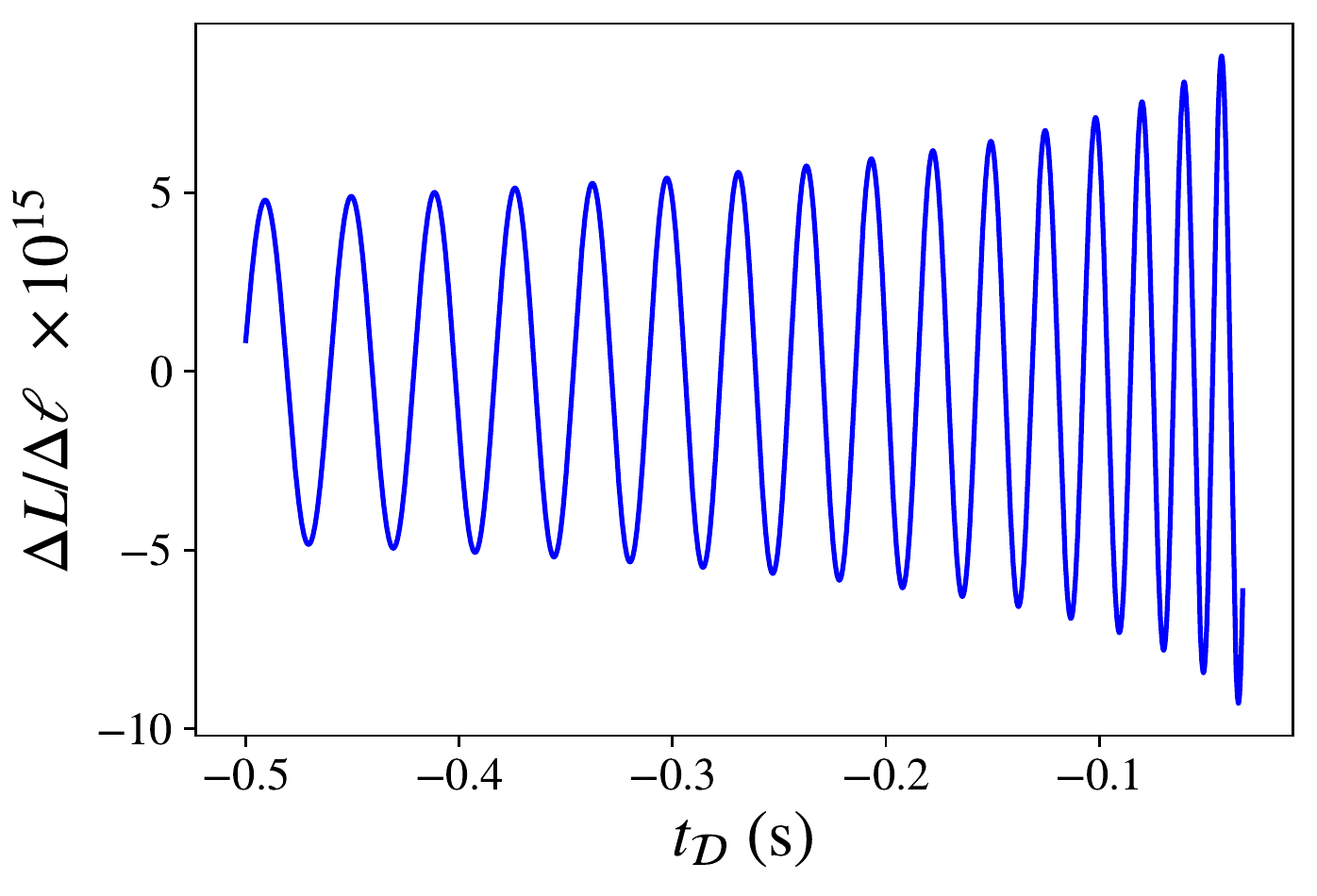}
	\caption{Frequency shift (left) and radar length perturbation (right) for a typical waveform of a binary system detected by aLIGO in the inspiral phase (up to a frequency $\omega_g \simeq$ 425 rad/s). We have made $\theta = \pi/2$, $\phi = \pi/3$, $\Delta \ell =$ 4 km, $M_c = 28.3 \, \text{M}_{\odot}$, $r = 410$ Mpc and $\Phi_0=t_c=\iota = 0$.}
\end{figure}

Finally, for illustrative purposes, figure \ref{fig:binary_waveform} shows the two key quantities discussed in this work, namely, light frequency shift in eq.~(\ref{redshift}) and the radar distance perturbation
\begin{align}
\Delta L \leqdef D_R(t_{\mathcal{D}} - \Delta \ell) - \Delta \ell
\end{align}
in eq.~(\ref{radardistspheric}), for a simple template of the GW amplitude emitted by a binary merger in its inspiral phase. It is given by \cite{Maggiore2007}
\begin{align}
h_+(t-x) &= \frac{1}{r} \left(\frac{GM_c}{c^2}\right)^{5/4} \left(\frac{5}{c \tau}\right)^{1/4} \frac{1 + \cos^2 \iota}{2} \cos[\Phi(\tau)]\,,\nonumber \\
h_{\times}(t-x) &= \frac{1}{r} \left(\frac{GM_c}{c^2}\right)^{5/4} \left(\frac{5}{c \tau}\right)^{1/4} \cos \iota \sin[\Phi(\tau)]\,,
\end{align}
where $M_c$ is the chirp mass of the binary system, $r$ is the luminosity distance from the source to the detector, $\iota$ is the angle between the normal direction of the binary plane and the line of sight and
\begin{equation}
\Phi(\tau) \leqdef -2 \left(\frac{5GM_c}{c^3} \right)^{-5/8} \tau^{5/8} + \Phi_0\,,
\end{equation}
with
\begin{equation}
\tau \leqdef t_{c} - t - x/c\,,
\end{equation}
where $\Phi_0$ is a constant and $t_{c}$ is the instant of coalescence of the merger as seen by an observer at $x = 0$. The values of $M_c$ and $r$ were chosen to match those of the first detected black hole binary system by aLIGO \cite{Abbott2016b}, while $\iota$, $\Phi_0$ and $t_{c}$ were set to vanish.

\subsection{Can frequency shift and arm length perturbation combine to cancel the interference pattern?} \label{subsec:freq_vs_radar_dist}

In interferometric experiments designed to detect GWs, one usually relies on the assumption that the detected non-trivial ($\bm{\epsilon}$-dependent) part of the final intensity pattern is, to linear order in $\bm{\epsilon}$, proportional to the phase difference between light beams after traveling their round-trip along each arm. If one assumes that the amplitude and polarization of the electromagnetic fields of the beams are not affected by the GW (cf. L2), this relation follows and is widely adopted in literature (e.g. \cite{Maggiore2007}). In fact, as we shall see in L2, there are corrections to it if the electromagnetic field is evolved according to its full propagation equation along null geodesics in curved spacetimes \cite{Santana2020}. For this subsection, we shall adopt this point of view to discuss one of the most common conceptual concerns arising as a consequence of this picture.

We know that when a GW reaches an interferometer, it changes the radar length of both arms in an anisotropic way, given by eq.~(\ref{radardistspheric}), implying the well known phase difference at the end of the process. The issue usually posed \cite{Saulson1997, Faraoni2007} is based on the following question: should not frequency shifts, as the one calculated in eq.~(\ref{redshift}), result in a contribution to the final phase difference, \emph{additionally} to that due to difference in round-trip travel time? After all, the change in light wavelength should heuristically result in a change of the distances between the electric field maxima. If this is true, could such contribution cancel the first in some specific case, so that no GW could be detected at all?

In the LIGO FAQ webpage \cite{LIGOFAQ}, the answer to the intimately related question ``If a gravitational wave stretches the distance between the LIGO mirrors, doesn't it also stretch the wavelength of the laser light?" begins with the following assertion: ``While it's true that a gravitational wave does stretch and squeeze the wavelength of the light in the arms \emph{ever so slightly}, it does NOT affect the fact that the beams will travel different distances as the wave changes each arm's length". But why is it that the only relevant quantity for the difference in phase is the difference in path traveled by light is never justified in the answer. 

Furthermore, at the end, the webpage concludes by saying that: ``(...) the wavelengths of light have no bearing on the all-important interference pattern. The effects of the length changes in the arms far outweigh any change in the wavelength of the laser, so we can virtually ignore it altogether.". Firstly, these effects, though minute, are of the same order in $\bm{\epsilon}$, preventing any \emph{a priori} conclusions about the negligibility of any of them when compared with the other. Although in the \emph{particular} case studied in figure \ref{fig:binary_waveform} the change in the arms' lengths indeed outweighs the change in the wavelength, this assertion does not seem straightforward for other regimes in the GW spectrum. In \cite{Faraoni2007}, Faraoni shows for the long-wavelength limit ($\lambdabar_g \gg \Delta\ell$, where $\lambdabar_g\leqdef \lambda_g/2\pi$ is the reduced GW wavelength) that, in fact, the frequency shift vanishes. In the frequency band where such a limit is valid for aLIGO, the conclusion in the webpage is then justified. But for the higher portion of the aLIGO detectable range of frequencies (kHz to $10$ kHz), such approximation does not seem to be acceptable (for kHz, $\lambdabar_g \approx 10$ km, while $\Delta\ell = 4$ km). Besides, for other GW detectors, such as LISA and the Cosmic Explorer, this approximation becomes even less applicable, since the former has $\Delta\ell = 2.5 \times 10^6$ km and will detect GWs in the range $10^{4} \text{\;km}< \lambdabar_g < 10^8 \text{\;km}$, and the latter has $\Delta\ell = 40$ km and will detect GWs in the range $5 \text{\;km}<\lambdabar_g<10^4 \text{\;km}$. Lastly, even if it were true that one effect is always dominant over the other, it is not obvious how do they contribute to the final intensity pattern so that one could be neglected when compared to the other (one of the main themes discussed in L2). Here we provide a simple answer to the problem raised regardless of any assumption on either the GW wavelength or the arm length, provided that the geometrical optics approximation for light is valid.    

In the electromagnetic geometrical optics regime, the phase of light  $\psi(x^{\mu})$ is related to the 4-dimensional wave vector by:
\begin{equation}
	k_{\mu} = \nabla_{\mu} \psi. \label{eq:phase}
\end{equation}
Since the frequency measured by the TT frame is $k_0$, the above equation suggests that $\psi$ is affected by how the quantity $\omega_{\textrm{e}}$ evolves along the light ray. However, the nullity of the light geodesic rays (\ref{pert_geod}) together with (\ref{eq:phase}) gives:
\begin{equation}
	k_{\mu}k^{\mu} = 0 \Rightarrow \frac{D \psi}{d \vartheta} = 0. \label{constphase}
\end{equation} 
This, in turn, implies that, although it is true that there is a frequency shift influence on the phase throughout the photon's round-trip, the changes in the spatial components $k_{i}$ or, in other words, the spatial trajectory perturbations, contribute to it as well, in such a way that the net effect is to preserve the constancy of $\psi$ throughout the null curve.

One can conclude from these considerations that the phase at the end of the round-trip of the beams in each arm is equal to its initial value, but, because the interferometer's arms are deformed distinctly, the rays that combine at the end must have been emitted in different events along $\mathcal{S}$, which, in general, leads them to have different values of $\psi$ and, consequently, to interfere in a non-trivial manner. Ultimately, the phase difference at the end occurs, indeed, solely because of the discrepant paths light travels in each arm. If one assumes this phase disparity to be the only contribution, to linear order in $\bm{\epsilon}$, to the intensity fluctuations, frequency shifts, although existent, should not explicitly stand as an independent or additional influence to be considered in the intensity measurements, but to be another manifestation of the change in radar distance, as illustrated by eq.~(\ref{Doppler}). Nonetheless, we shall demonstrate in L2 that if the electric field magnitude is evolved using its full propagation equation \cite{Santana2020}, a frequency shift contribution does arise from the perturbation on each photon's energy. 
\section{Conclusion}
\label{sec:conclusion}

In this first part of the two article series on the influence of gravitational waves upon light we have:
\begin{enumerate}[(i)]
	\item showed that the round-trip travel time of light between two observers of the TT frame in a GW spacetime whose background is flat does not change if spatial trajectory perturbations in the luminous path due to GWs are neglected.

	\item elucidated the necessary and sufficient conditions for the above result to be valid and on which observables the simplified spatial trajectory scheme could give wrong results.
	
	\item obtained explicitly, in terms of known parameters, the family of possible parametrized null geodesic arcs exchanged by two TT observers.
	
	\item calculated the light frequency shift, relating it with the radar distance time change.
	
	\item used the laws of geometrical optics to explain how interferometers can detect GWs even though frequency shifts are present.
\end{enumerate}

Furthermore, we have prepared the ground for the discussion of the electric field propagation in a toy model interferometric experiment as a way of determining the final interference pattern, which will be the theme of the second part of this study (L2). This was done by the introduction, calculation and discussion of two elements that will play a major role on such propagation: the radar distance and the frequency shift of light. Using the newly found equation for the electric field propagation for general spacetimes and reference frames derived in \cite{Santana2020}, we will show how those two quantities appear in the final interference pattern and a comparison on their significance will be made as a way to justify the common picture of such pattern depending only on the phase difference of the two interacting light beams.
  
\acknowledgments
ISM thanks Brazilian funding agency CNPq for PhD scholarship GD 140324/2018-6 and JCL thanks Brazilian funding agencies CAPES and FAPERJ for MSc scholarships 31001017002-M0 and 2016.00763-4, respectively.

\appendix

\section{On the domains of the $\epsilon$-parametrized light rays}
\label{app:domains}

Each one of the outgoing null geodesic arcs from eq.~(\ref{consistent_models}) is a function
\begin{align}
\xi_{(\bm{\bm{\epsilon}})}: & \, D_{(\bm{\bm{\epsilon}})} \rightarrow \mathbb{R}^4\,,
\end{align}
whose domain is
\begin{align}
D_{(\bm{\epsilon})} \leqdef [0, \vartheta_{\mathcal{R}}]\,, \label{domains}
\end{align}
and which satisfies the discussed mixed conditions related to eq.~(\ref{mixed_conditions}). More precisely, thinking about the parametrized curves as functions of $(\bm{\bm{\epsilon}}, \vartheta, \bm{P})$, one can evaluate $\xi_{(\bm{\bm{\epsilon}})}^i$ at its final event, substituting the spatial coordinates of $\mathcal{R}$, $x^i_{\mathcal{M}} \in \bm{P}$, and then solving for the value $\vartheta_{\mathcal{R}}$:
\begin{align}
\xi_{(\bm{\bm{\epsilon}})}^i(\vartheta_{\mathcal{R}}) = x^i_{\mathcal{M}} \; \Rightarrow \; \vartheta_{\mathcal{R}} = f(\bm{\bm{\epsilon}}, \bm{P})\,. \label{vartheta}
\end{align}
So $\mathcal{R}$ and, consequently, $\vartheta_{\mathcal{R}}$ are determined by both the parameters $\bm{P}$ and $\bm{\bm{\epsilon}}$. This is why $D_{(\bm{\bm{\epsilon}})}$ depends on $\bm{\bm{\epsilon}}$. Of course, since eq.~(\ref{domains}) holds for all $\bm{\epsilon}$ including $\bm{0}$, the first term of the expansion, $f(\bm{0}, \bm{P})$, will be equal to what we call $\vartheta_{\mathcal{R}_{(0)}}$ in the main text.

Also, for all $\vartheta \in D_{(\bm{\epsilon})} \cap D_{(\bm{0})}$, it is true that
\begin{align}
\xi_{(\bm{\epsilon})}(\vartheta) = \xi_{(\bm{0})}(\vartheta) + \epsilon_P\,\xi^P(\vartheta)\,, \label{param_curve}
\end{align}
from which one derives eq.~(\ref{k_expansion}). The above expansion consists on splitting the functional dependence of the curve into that of the model $\mathbb{M}_{(\bm{0})}$ plus some additional terms.

Of course we can still write eq.~(\ref{param_curve}) for all points in $D_{(\bm{\epsilon})}$ as long as we extend the unperturbed geodesic arc to $\xi_{(\bm{0})|\textrm{ext}}$, maintaining its geodesic character, to this domain (or even to $\cup_{\bm{\epsilon}} D_{(\bm{\epsilon})}$). However, in this case, one should be aware that, while evaluating eq.~(\ref{param_curve}) in $\vartheta_{\mathcal{R}}$, the first term, although written with a subscript $(\bm{0})$, will have contributions depending on $\bm{\epsilon}$, since, from eq.~(\ref{vartheta}):
\begin{align}
\xi_{(\bm{0})|_{\textrm{ext}}}(\vartheta_{\mathcal{R}}) &= \xi_{(\bm{0})}(\vartheta_{\mathcal{R}_{(\bm{0})}}) + k_{(\bm{0})}(\vartheta_{\mathcal{R}_{(\bm{0})}})\frac{\partial f}{\partial \epsilon_P}(\bm{0}, \bm{P}) \epsilon_P\, \nonumber \\
&= \xi_{(\bm{0})}(\vartheta_{\mathcal{R}_{(\bm{0})}}) + \sum_P \epsilon_P\Delta_{\epsilon_P}(\xi_{(\bm{0})|\textrm{ext}}(\vartheta_{\mathcal{R}}))\,.
\end{align}
This illustrates how $\Delta_{\epsilon_P}$ appearing in the discussion of section \ref{subsec:paramet_equations} (there, for a single parameter $\epsilon$) can be perceived either as a variation in the form or as a total variation, depending on which function it acts upon. For instance:
\begin{align}
 \Delta_{\epsilon_P}\left(\xi_{(\bm{\epsilon})}\circ f\right) \neq \Delta_{\epsilon_P}\left(\xi_{(\bm{\epsilon})}\right)\circ f\,.
\end{align}

\bibliographystyle{unsrt}
\bibliography{interaction_EM_waves_and_GW_part_1}

\end{document}